\documentclass[%
reprint,
superscriptaddress,
nofootinbib,
 amsmath,amssymb,
 aps,
prl,
]{revtex4-2}

\usepackage{graphicx}% Include figure files
\usepackage{dcolumn}% Align table columns on decimal point
\usepackage{bm}% bold math
\usepackage{xcolor}
\usepackage{booktabs}
\usepackage{subcaption} % Required for subfigures  
\usepackage{float}

\begin{document}

\preprint{APS/123-QED}

\title{Ultrafast dynamics and light-induced superconductivity from first principles}

\affiliation{%
 Research Laboratory of Electronics, Massachusetts Institute of Technology \\
 50 Vassar Street, Cambridge, MA, USA, 02139-4307
}%

\affiliation{%
 Institute of Theoretical and Computational Physics, Graz University of Technology \\ 
 Petersgasse 16, 8010 Graz, Austria
}%

\affiliation{%
 Enterprise Science Fund, Intellectual Ventures \\ 
 Bellevue, Washington, United States
}%

\author{Alejandro Simon}
\email{alejansi@mit.edu}

\affiliation{%
 Research Laboratory of Electronics, Massachusetts Institute of Technology \\
 50 Vassar Street, Cambridge, MA, USA, 02139-4307
}%

\author{James Shi}

\affiliation{%
 Research Laboratory of Electronics, Massachusetts Institute of Technology \\
 50 Vassar Street, Cambridge, MA, USA, 02139-4307
}%

\author{Eva Kogler}
\affiliation{%
 Institute of Theoretical and Computational Physics, Graz University of Technology \\ 
 Petersgasse 16, 8010 Graz, Austria
}% 

\author{Reed Foster}

\affiliation{%
 Research Laboratory of Electronics, Massachusetts Institute of Technology \\
 50 Vassar Street, Cambridge, MA, USA, 02139-4307
}%

\author{Dominik Spath}
\affiliation{%
 Institute of Theoretical and Computational Physics, Graz University of Technology \\ 
 Petersgasse 16, 8010 Graz, Austria
}%

\author{Emma Batson}
\affiliation{%
 Research Laboratory of Electronics, Massachusetts Institute of Technology \\
 50 Vassar Street, Cambridge, MA, USA, 02139-4307
}%

\author{Pedro N. Ferreira}
\affiliation{%
 Institute of Theoretical and Computational Physics, Graz University of Technology \\ 
 Petersgasse 16, 8010 Graz, Austria
}%

\author{Mihir Sahoo}
\affiliation{%
 Institute of Theoretical and Computational Physics, Graz University of Technology \\ 
 Petersgasse 16, 8010 Graz, Austria
}%

\author{Rohit Prasankumar}
\affiliation{%
 Deep Science Fund, Intellectual Ventures \\ 
 Intellectual Ventures, Bellevue, Washington, United States
}%

\author{Phillip D. Keathley}
\affiliation{%
 Research Laboratory of Electronics, Massachusetts Institute of Technology \\
 50 Vassar Street, Cambridge, MA, USA, 02139-4307
}%

\author{Karl K. Berggren}
\affiliation{%
 Research Laboratory of Electronics, Massachusetts Institute of Technology \\
 50 Vassar Street, Cambridge, MA, USA, 02139-4307
}%

\author{Christoph Heil}
\affiliation{%
 Institute of Theoretical and Computational Physics, Graz University of Technology \\ 
Petersgasse 16, 8010 Graz, Austria
}%

\date{\today}% It is always \today, today,
             
\begin{abstract}
Experiments on superconducting materials have unveiled unique emergent properties when they are driven far from equilibrium. However, a quantitative first-principles treatment that describes experimental observations is lacking. In this work, we develop an \textit{ab-initio} model for the nonequilibrium response of optically irradiated superconducting films within the framework of conventional electron-phonon-mediated superconductivity, leveraging new numerical techniques to solve the Migdal-Eliashberg equations directly on the real-frequency axis. This enables us to quantitatively reproduce the optical response of superconducting films in pump-probe experiments and validate our approach on measurements of the differential reflectance of Pb and LaH$_{10}$ in response to a pump excitation. Similar calculations performed on the alkali-doped fulleride K$_3$C$_{60}$ reveal that a photo-induced superconducting state is generated after irradiation by an ultrafast mid-infrared pulse of sufficient intensity, as reported in prior experimental work. The enhancement in this framework is attributed to the excitation of quasiparticles to energies resonant with the strongest electron-phonon coupling in K$_3$C$_{60}$, in close analogy to the mechanism for enhancement of superconductivity under microwave irradiation, explaining the nature of the photo-induced superconducting state and elucidating the subsequent quasiparticle and phonon dynamics. Our results suggest that photo-induced superconductivity is accessible in more materials than previously recognized. We demonstrate this by performing calculations on calcium-intercalated graphite, CaC$_6$, and predict a similar photo-induced superconducting gap. 

\end{abstract}

\maketitle

Collective excitations induced by driving a solid far from equilibrium can give rise to unique emergent behaviour \cite{basov2017towards, de2021colloquium}, including Floquet-engineered phases \cite{floquet}, nonlinear plasmonic responses \cite{Kauranen2012-qr}, and coherently driven phonon, spin, or charge-order dynamics \cite{Mitrano2016-pk}. The study of these excitations has been driven forward by the development of ultrafast lasers that are capable of producing pulses several orders of magnitude shorter than the excitation lifetimes \cite{Guo2024-ve}. In particular, bound states, such as Cooper pairs, excitons, and polaritons, which exhibit relatively long lifetimes in comparison with the pulse widths achievable with modern laser technology, are ideal platforms to study and search for novel fundamental physics.

\begin{figure*}
    \centering
    \includegraphics[width=1\linewidth]{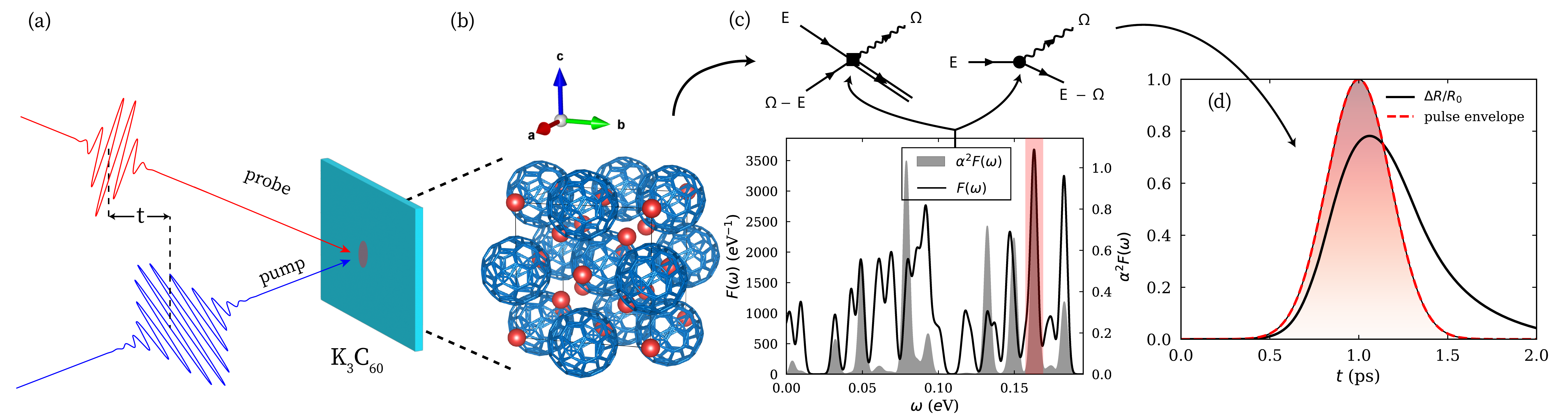}
    \caption{Modeling nonequilibrium dynamics in a superconductor from first principles. (a) Schematic of a pump-probe experimental set-up. A high-energy pump creates a local nonequilibrium excitation in the sample. After a time delay $\tau$ a probe pulse is then used to measure the optical properties of the sample. The nonequilibrium response is then mapped by varying $\tau$. (b) Crystal structure of K$_3$C$_{60}$ used to compute (c) the quasiparticle and phonon interactions that depend on the Eliashberg spectral function $\alpha^2F(\omega)$ and phonon density of states $F(\omega)$. The red bar denotes the $170\,$meV peak in the electron-phonon coupling of K$_3$C$_{60}$ that was pumped in the experiments in Refs. \cite{Mitrano2016-pk, budden2021evidence, Rowe2023-ab}. The quantities $\alpha^2F(\omega)$ and $F(\omega)$ are used to compute the transition probabilities for nonequilibrium excitations. (d) These interactions are then used to compute the dynamic response of the material, such as the differential reflectance $\Delta R/R_0$ shown in black in response to a pulse envelope shown in red.}
    \label{fig:pump-probe}
\end{figure*}

A wide array of literature has focused on experimentally studying excitations of these bound states, usually unveiling fascinating and unusual dynamics \cite{Mitrano2016-pk}. Studies of Cooper pair and Bogoliubov quasiparticle excitations in superconductors pose a particularly interesting case, as the observed collective excitations have both deeply fundamental and practical implications. For example, Higgs oscillations \cite{PhysRevB.92.224517, higgs-mode-review}, photo-induced superconductivity \cite{fausti2011light, Mitrano2016-pk, budden2021evidence, von2022amplification, Rowe2023-ab}, and superconducting detectors \cite{2001ApPhL..79..705G, simon2025abinitiomodelingnonequilibrium} provide some examples of how driving superconductors out of equilibrium can reveal fundamental collective dynamics and enable technologically relevant functionalities. Furthermore, the phenomenon of photo-induced superconductivity, which has been observed in organic conductors \cite{Mitrano2016-pk, PhysRevX.10.031028, PhysRevLett.127.197002, budden2021evidence, Rowe2023-ab} and in the unconventional superconducting cuprates \cite{fausti2011light, von2022amplification}, remains an exciting prospect for demonstrating superconductivity near room-temperature \cite{doi:10.1073/pnas.2520324123}. 

Significant theoretical progress has been made in describing these excitations \cite{Komnik-2016, PhysRevB.94.214504, PhysRevB.94.155152, PhysRevB.93.144506, doi:10.1126/sciadv.1700718, Kennes2017-qo, PhysRevB.98.235149, Nava2018-ky, chattopadhyay2026giant}. However, a fully self-consistent first-principles description of optical irradiation in superconductors capable of making quantitative predictions for arbitrary realistic materials remains challenging and typically relies on substantial approximations. Consequently, theoretical predictions lag experimental realizations. 

In this letter, we close this gap by developing a predictive first-principles theory of optical excitations in conventional superconductors. We demonstrate that our approach quantitatively reproduces pump–probe experiments in two materials with vastly different electron–phonon coupling strengths: Pb and LaH$_{10}$. Beyond quantitative agreement with experiment, our framework provides a unified and physically transparent description of nonequilibrium dynamics. In particular, we show that it naturally explains both the transient enhancement of the superconducting gap under optical irradiation and the photo-induced superconducting response observed in K$_3$C$_{60}$. This analysis reveals how quasiparticle and phonon dynamics following irradiation govern the characteristic experimental signatures and suggests that these processes are crucial to consider for achieving metastability of the photo-induced state. Finally, guided by these insights, we identify calcium-intercalated graphite, CaC$_6$, as a promising candidate material that should exhibit a photo-response similar to K$_3$C$_{60}$.

In pump-probe spectroscopy of superconductors, the superconductor is illuminated by an optical pump pulse, generating an excitation in the condensate that is examined by measuring the optical properties of the film with a probe sent after a delay $\tau$. This process is schematically depicted in Figure \ref{fig:pump-probe}a. With modern equipment, femtosecond time resolution can be achieved, which can thus provide information about the underlying microscopic processes, as dynamics are expected to occur on timescales of $\tau_\Delta \sim\hbar/\Delta_0 \lesssim \rm{ps}$, where $2\Delta_0$ is the zero-frequency gap, while thermalization takes much longer \cite{simon2025abinitiomodelingnonequilibrium}. Quasiparticle relaxation following absorption of the pump occurs through a mixture of electron-electron scattering, quasiparticle scattering and recombination, and phonon pair-breaking and scattering \cite{chang1977kinetic, simon2025abinitiomodelingnonequilibrium, simon2025ab}. Depending on the excitation, material, and temperature, these processes can last from a few picoseconds to nanoseconds \cite{kaplan1976quasiparticle}.

These dynamics can be treated naturally with the Green's function formalism. In principle, as long as the electron and phonon Green's functions and self-energies can be computed, the transport properties of a material can be obtained with the Kadanoff-Baym equations \cite{kadanoff2018quantum, prange1964transport}. When the quasiparticle approximation holds, we can reduce the Kadanoff-Baym transport equations to a set of kinetic equations for the quasiparticle distribution $f(E)$ and phonon distribution $n(\omega)$. The consequences of this reduction are detailed in the SI \cite{seeSI}, and the resulting nonlinear differential equations to solve for the quasiparticle $f(E)$ and phonon $n(\omega)$ distribution functions are then
\begin{subequations}
\begin{equation}
\label{eq:f}
\frac{df(E)}{dt} = I_{e\mathrm{-ph}}(E) + I_{\mathrm{ph-}e}(E) + I_{\mathrm{R}}(E)
\end{equation}
\begin{equation}
\label{eq:n}
    \frac{dn(\omega)}{dt} = I_{\mathrm{S}}(\omega) + I_{\mathrm{B}}(\omega) - \frac{n(\omega) - n^{\mathrm{eq}}(\omega)}{\tau_{\mathrm{esc}}},
\end{equation}
\end{subequations}
where the collision integrals $I_{e\mathrm{-ph}}(E)$, $I_{\mathrm{ph-}e}(E)$, $I_{\mathrm{R}}(E)$, $I_{\mathrm{S}}(\omega)$, and $I_{\mathrm{B}}(\omega)$ correspond to electron-phonon scattering, quasiparticle recombination, phonon scattering, and phonon pair-breaking, respectively, and are defined in the SI \cite{seeSI}. Embedded within the collision integrals and central to determining the transition probabilities are the Eliashberg spectral function $\alpha^2 F(\omega)$ and phonon density of states $F(\omega)$; $n^{\mathrm{eq}}(\omega)$ is the equilibrium Bose distribution, and $\tau_{\rm{esc}}$ is the characteristic phonon escape time included to provide a dissipative channel and prevent runaway heating. The quantity $\tau_{\mathrm{esc}}$ can be estimated from the acoustic mismatch model~\cite{little1959transport}. 

A full treatment within the framework of conventional electron-phonon-mediated superconductivity requires these kinetic equations to be solved self-consistently with the Migdal-Eliashberg equations on the real-frequency axis. While this computation would ordinarily be prohibitively expensive, at the core of our approach is an efficient real-frequency-axis solver for the Migdal-Eliashberg equations \cite{FIXME-THIS-PAPER}, which is a departure from the traditional procedure of solving on the imaginary-frequency axis \cite{IsoME, ponce2016epw}. Solving directly on the real-frequency axis has two key advantages: it provides immediate access to experimentally relevant quantities such as the quasiparticle density of states and optical properties, circumventing the need for the ill-conditioned analytic continuation from imaginary to real frequencies \cite{kraberger2017maximum, PhysRevB.61.5147, khodachenko2024nevanlinna}; and it naturally enables the computation of nonequilibrium dynamics. Within the constant electronic density of states approximation, our implementation solves the Migdal-Eliashberg equations in under a second on a standard laptop \cite{FIXME-THIS-PAPER}.

To model the pump-probe excitation, we add to the right-hand side of Eq. \eqref{eq:f} a pump term 
\begin{align}
\label{eq:pump}
I_{\rm{pump}}(\varepsilon, t)& = \Gamma_{\mathrm{pump}} S(t) \int d\varepsilon'  \rho(\varepsilon',t) \nonumber \Big| M(\varepsilon, \varepsilon', t) \Big|^2 \nonumber \\
& \times \Bigg\{ [1-f(\varepsilon,t)]f(\varepsilon',t) \delta(\varepsilon - \varepsilon' - \omega_{\mathrm{pump}}) \nonumber \\
& - f(\varepsilon,t)[1-f(\varepsilon',t)] \delta(\varepsilon - \varepsilon' + \omega_{\mathrm{pump}}) \Bigg\}
\end{align}
where $\Gamma_{\mathrm{pump}}$ is fixed by matching the absorbed energy to the deposited energy, $S(t) = \rm{exp}[-4 \ln 2 (t/\tau_{\mathrm{pump}})^2]$, $\tau_{\mathrm{pump}}$ is the pulse envelope, $\omega_{\mathrm{pump}}$ is the pump photon energy, $\rho(\varepsilon,t)$ is the normalized quasiparticle density of states, and $M(\varepsilon,\varepsilon',t)$ are the usual coherence factors. For excitation energies $\omega_{\mathrm{pump}} \gg 2\Delta$, additional electron-electron and pair-breaking interactions should be included. However, this is only relevant for the first few $\sim$fs, and the longer $\sim$ps time scale in which we are interested in is dominated by electron-phonon interaction and relaxation. 

\begin{figure}
   \centering
   \includegraphics[width=\linewidth]{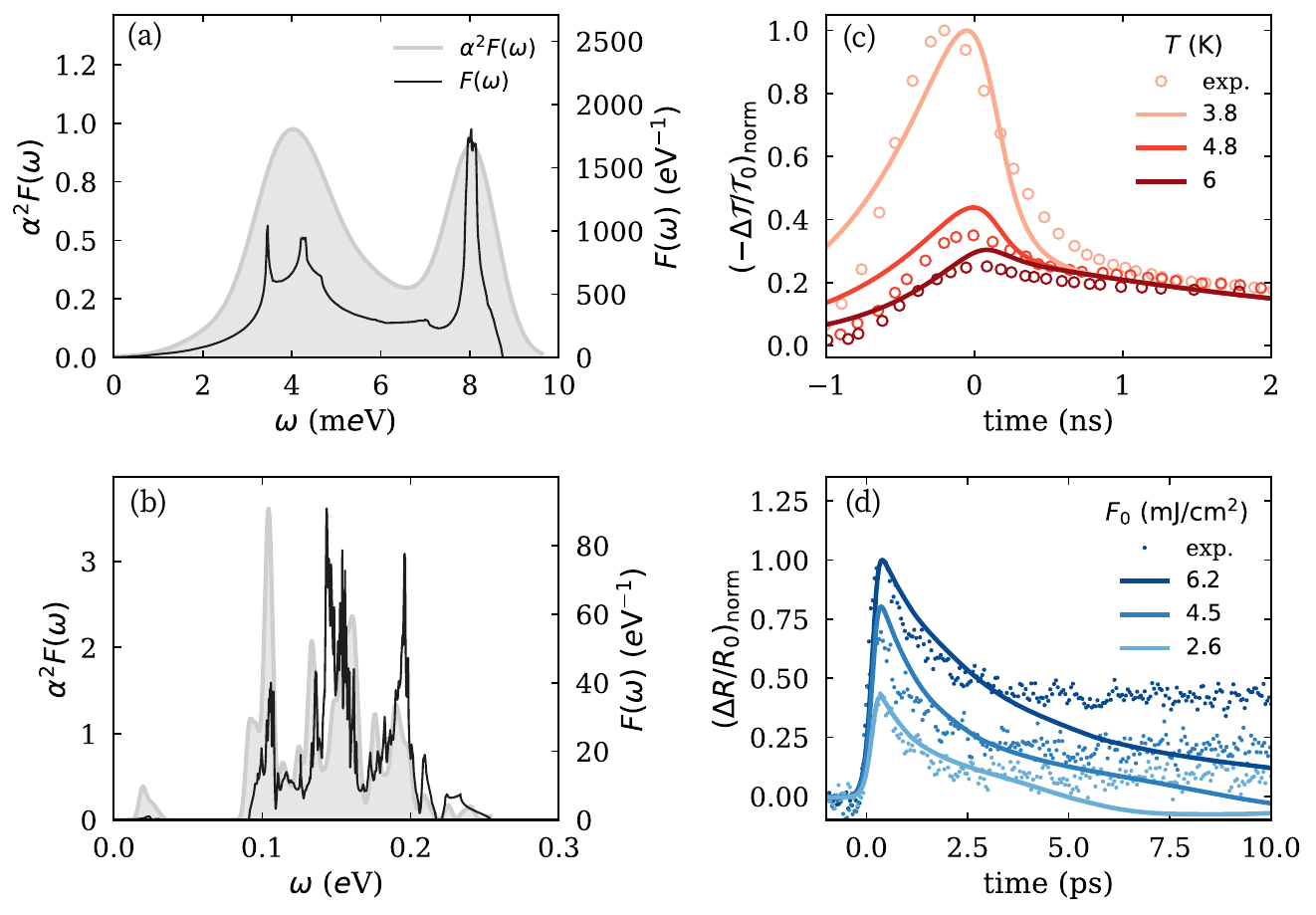}
    \caption{Calculated optical response of superconducting films to excitation from a pump pulse. (a-b) The Eliashberg spectral function $\alpha^2F(\omega)$ and $F(\omega)$ for (a) Pb and (b) LaH$_{10}$. (c) Negative differential transmission $-\Delta \mathcal{T}/\mathcal{T}_0$ for Pb at various temperatures as a function of time for a pump frequency of 1.5$\,e$V and probe frequency of $5\,\mathrm{m}e\mathrm{V}$ with experimental data manually extracted from Ref. \cite{PhysRevB.72.024510}. (d) Differential reflectance $\Delta R/R_0$ for LaH$_{10}$ at $T=77\,$K for a pump frequency of 2$\,e$V and probe frequency of 1.5$\,e$V and various fluences $F_0$ with experimental data from Ref. \cite{Wu2024-rm}. Here, the simulated and experimental $-\Delta\mathcal{T}/\mathcal{T}_0$ and $\Delta R/R_0$ are normalized to the maximum of the displayed data.}
   \label{fig:results1}
\end{figure}

With the solutions for the full electronic Green's function, we compute the optical properties of the material. From linear-response theory, complex conductivities $\sigma(\omega,t)$ are calculated by solving a Kubo formula \cite{mahan2013many, nam1967theory} and the refractive index is then obtained from the dielectric function $n(z, \omega, t) = \sqrt{\epsilon(z, \omega,t)} = \sqrt{1 + i \sigma(z, \omega,t)/\epsilon_0\omega}$, where $\epsilon_0$ is the permittivity of free-space. When the pump interacts with the material, a change in $n(z, \omega,t) = n_0(\omega) + \Delta n(z, \omega,t)$ is produced, where $n_0(\omega)$ is the equilibrium index of the material. Depending on the pump penetration depth, this disturbance can be nonuniform across the depth $z$ of the material. The measured reflectance $R$ of the sample at the probe frequency then depends on the details of the sample geometry and its interaction with the pump. A careful treatment of this detail, including the nonequilibrium probe penetration depth, is important to avoid potentially large errors \cite{PhysRevLett.130.146002}. We address this in greater detail in the SI \cite{seeSI}. The most general approach is to solve Eqs. \eqref{eq:f} and \eqref{eq:n} along the full thickness of the material while accounting for the nonlinear deposition of the pump energy with thickness, and then solving the Maxwell equations to obtain $R(\omega,t)$ and $T(\omega,t)$. With $R(\omega,t)$, we obtain the differential reflectance
\begin{equation}
    \frac{\Delta R (\omega,t)}{R_0} = \frac{R(\omega, t) - R(\omega, t=-\infty)}{R(\omega, t=-\infty)}.
\end{equation} To fix $\Gamma_{\mathrm{pump}}$ we impose self-consistency between the absorbed energy density $E_{\mathrm{abs}} = F_0 \, A \, \chi(\omega)$ and the deposited energy density $E_{\mathrm{dep}}$, where $F_0$ is the incident fluence, $\chi(\omega) = 1 - R(\omega) - T(\omega)$ is the absorbed fraction, and $A$ is the beam area.

To solve Eqs. \eqref{eq:f} and \eqref{eq:n}, we compute $\alpha^2F(\omega)$ and $F(\omega)$ with density functional (perturbation) theory (DF(P)T)~\cite{giustino2017electron, QE-2017, Box2023-fhiaims-fric}. We apply this framework to two canonical superconductors spanning vastly different phonon energy scales: Pb at ambient pressure, with its low-energy phonon spectrum, and LaH10$_{10}$ at 165$\,$GPa, where superconductivity is driven by high-energy hydrogen vibrations. For Pb, shown in Figure \ref{fig:results1}a, we include spin–orbit coupling \cite{simon2025abinitiomodelingnonequilibrium}; for LaH10$_{10}$, shown in Figure \ref{fig:results1}b, quantum ionic effects and anharmonicity are explicitly accounted for using the stochastic self-consistent harmonic approximation (SSCHA) \cite{Bianco2017-SSCHA1, Monacelli2018-SSCHA2, Monacelli2021-SSCHA3}. The results for K$_3$C$_{60}$ are presented in Figure \ref{fig:pump-probe}c, and CaC$_6$ is shown in the SI \cite{seeSI}.

\begin{figure}[t]
    \centering
    \includegraphics[width=0.9\linewidth]{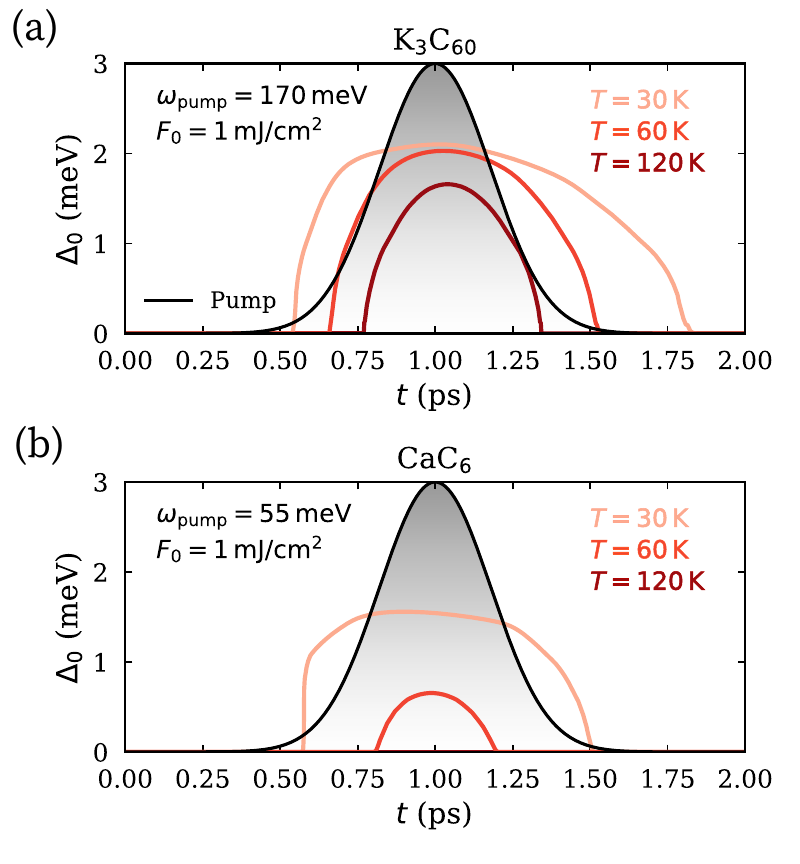}
    \caption{The photo-induced gap $\Delta_0$ in (a) K$_3$C$_{60}$ and (b) CaC$_6$ in response to excitation by mid-infrared $170\,$m$e$V and $57\,\mathrm{m}e\mathrm{V}$ pump pulses, respectively, with the pump pulse envelope shown in shaded grey for reference. The enhancement in $\Delta_0$ is shown for various temperatures above $T_{\mathrm{c0}}=15\,\mathrm{K}$ for K$_3$C$_{60}$ and $T_{\mathrm{c0}}=11\,\mathrm{K}$ for CaC$_{6}$.}
    \label{fig:K3C60}
\end{figure}

\begin{figure}[t]
    \centering
    \includegraphics[width=0.9\linewidth]{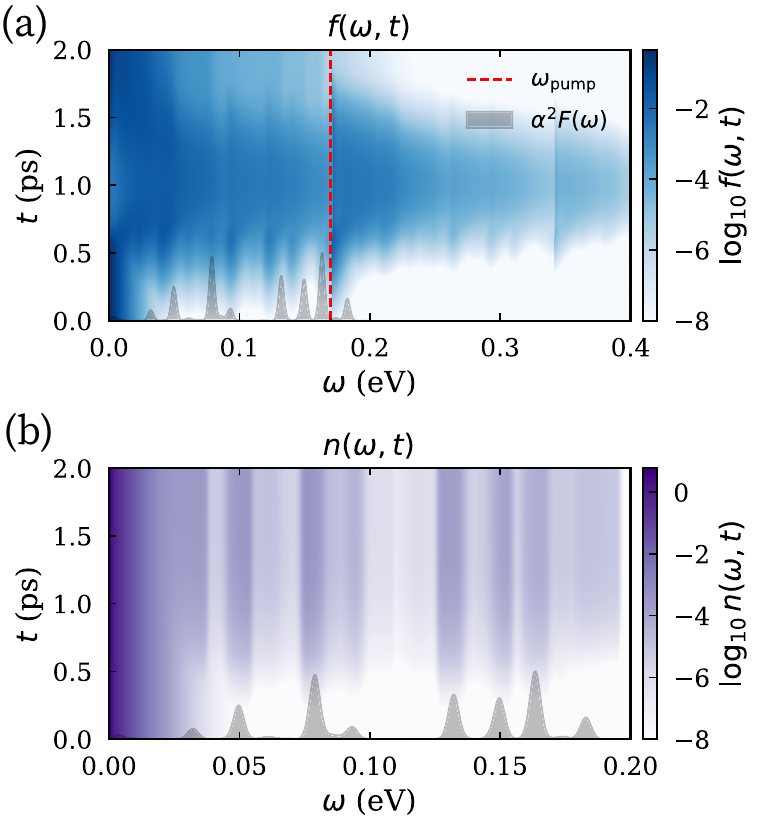}
    \caption{The nonequilibrium (a) quasiparticle $f(\omega,t)$ and (b) phonon $n(\omega,t)$ distribution functions for K$_3$C$_{60}$ at $T=30\,$K. The pump frequency $\omega_{\mathrm{pump}}=170\,$m$e$V is shown in (a) to emphasize the discontinuity in the distribution at $\omega_{\mathrm{pump}}$. Peaks in $f(\omega,t)$ and $n(\omega,t)$ appear corresponding to the structure in $\alpha^2F(\omega)$, with $\alpha^2F(\omega)$ plotted as a visual reference in grey.}
    \label{fig:f_n}
\end{figure}

With this framework in place, we investigate from first principles the dynamic response of a superconducting film to photoexcitation, and compare with experiment. In Figure \ref{fig:results1}c, we show the change in the negative differential transmittance $-\Delta \mathcal{T}/\mathcal{T}_0$ of Pb for a probe frequency $\omega_{\mathrm{pr}}$ in the far-infrared, computed for $\omega_{\mathrm{pump}} = 1.5\,e\mathrm{V}$, $\tau_{\mathrm{pump}}=2\,$ps, and $F_0\approx5\times10^{-7}\,$mJ/cm$^2$ to match the experimental conditions of Ref.~\cite{PhysRevB.72.024510}. Here, we use a reasonable estimate of $\tau_{\mathrm{esc}} = 500\,$ps to account for the poor cooling power of the apparatus and convolve the calculated response of $\Delta \mathcal {T}/\mathcal{T}$ with a $200\,$ps gaussian pulse to account for the long probe pulse \cite{PhysRevB.72.024510}. The Pb sample is also assumed to be in the thin-film limit $d \ll \delta(\omega) < \lambda$ \cite{PhysRevB.72.024510}, where $d$ is the sample thickness, $\delta(\omega) = 1/\alpha(\omega)$ is the penetration depth, $\alpha(\omega)= 2\omega/c\, \Im n(\omega)$ is the absorption coefficient, and $\lambda$ is the photon wavelength. Thus, $R(\omega,t)$ and $T(\omega,t)$ are obtained from the Fresnel equations \cite{PhysRevB.54.700, PhysRevB.72.024510}. The temperature dependence of the recovery time and amplitude of the response in $-\Delta \mathcal{T}/\mathcal{T}_0$ quantitatively agrees very well with the experimental data presented in Ref. \cite{PhysRevB.72.024510}. As the temperature approaches $T_{\rm{c}}$, the quasiparticle lifetime decreases \cite{kaplan1976quasiparticle}, leading to a faster relaxation of $\Delta(\omega,t)$ to equilibrium.  

In Figure \ref{fig:results1}d, we show the result of a similar calculation for LaH$_{10}$, with $\omega_{\mathrm{pump}}=2\,e$V and $\omega_{\mathrm{pr}}=1.5\,e$V with $\tau_{\mathrm{pump}}=70\,$fs and $F_0=2.6\,$mJ$/$cm$^2$, 4.5$\,$mJ$/$cm$^2$, and 6.2$\,$mJ$/$cm$^2$, consistent with the experimental apparatus in Ref. \cite{Wu2024-rm}. Here, the sample is not in the thin-film limit; however, given the similar pump and probe frequencies and the short ($\sim20\,\mathrm{nm}$) penetration depth in LaH$_{10}$, the excitation and probe are localized near the surface, varying at most by 3$\,$nm for the largest $\Delta n$ at $F_0 = 6.2\,$mJ/cm$^2$ \cite{seeSI}. Thus, we can treat the sample as a single layer with index $n(\omega,t)$ and calculate $R(\omega,t)$ with the Fresnel equation. Under this approximation, the phenomenological phonon escape term describes phonon escape for the pair-breaking phonon population within the optically weighted region, which is localized near the surface in LaH$_{10}$. As this cannot adequately capture the true phonon propagation dynamics in the bulk, the result is an underestimation of the saturation caused by the phonon bottleneck effect \cite{seeSI, Wu2024-rm}. Nevertheless, there is strong qualitative agreement between the theoretically predicted response in $\Delta R/R_0$ and the measurement. In particular, our model captures the presence of multiple timescales associated with the down-conversion of quasiparticle energy to phonons, phonon dissipation, and a phonon bottleneck that becomes more pronounced at larger $F_0$. The agreement in conventional superconductors with significantly different forms of $\alpha^2F(\omega)$ (Figs. \ref{fig:results1}(a,b))  confirms the robust nature of our model. To our knowledge, the framework we present here constitutes the first within the context of the Migdal-Eliashberg theory that is capable of quantitatively modeling and predicting the nonequilibrium properties of superconducting materials at arbitrary temperatures.

Motivated by the strong agreement between our calculations and experiment, we turn our attention to the alkali-doped fulleride K$_3$C$_{60}$, a material in which potential light-induced superconductivity has been observed. Going beyond previous extensive theoretical work \cite{Komnik-2016, PhysRevB.94.214504, PhysRevB.94.155152, PhysRevB.93.144506, doi:10.1126/sciadv.1700718, Kennes2017-qo, PhysRevB.98.235149, Nava2018-ky, chattopadhyay2026giant}, here we provide a framework that quantitatively estimates the transient enhancement in superconductivity under photo-excitation, demonstrates its connection to conventional electron-phonon dynamics, and describes the subsequent quasiparticle dynamics. In Ref. \cite{Mitrano2016-pk}, the large enhancement in $\Delta_0=\Delta(\omega = 0)$ was attributed to a potential resonance and excitation of the $\sim$170$\,$m$e$V vibrational mode, which exhibited a large peak in their calculated $\alpha^2F(\omega)$. We have performed phonon and electron-phonon interaction calculations using FHI-aims \cite{Blum2009-fhiaims, Box2023-fhiaims-fric}, revealing a similarly pronounced peak in $\alpha^2F(\omega)$, as shown in Fig. \ref{fig:pump-probe}c.

Remarkably, in replicating the experimental parameters presented in Ref. \cite{Mitrano2016-pk}, we obtain a photo-induced gap $\Delta_0(t)$ at $T > T_{\mathrm{c0}}$, corresponding to a Cooper instability above $T_{\mathrm{c0}}$, where $T_{\mathrm{c0}}=15\,\mathrm{K}$ is the equilibrium critical temperature for K$_3$C$_{60}$ estimated from DFT. As shown in Figure \ref{fig:K3C60}a, this effect persists at temperatures much larger than $T_\mathrm{c0}$ and for a short time after the pump is turned off. The same calculation was repeated for CaC$_6$, revealing a similar response to optical irradiation as depicted in Figure \ref{fig:K3C60}b. We attribute this photo-induced gap to the excitation of quasiparticles to energies resonant with $\alpha^2F(\omega)$, where the pairing interaction is strongest, leading to a concomitant depletion of quasiparticles through pairing and establishing a superconducting channel with $\Delta_0 > 0$. The increase in $\Delta_0$ therefore has a short-lived transient nature, which depends on the details of the electron-electron and electron-phonon relaxation in the film, with the photo-induced $\Delta_0$ vanishing as the distribution thermalizes in time $\tau_{\mathrm{th}}$. The relevant time scale for observing this state is then $\tau_{\mathrm{th}} > \hbar/\Delta_0$. 

This proposed mechanism is analogous to the enhancement of $\Delta_0$ under continuous microwave irradiation, where excitation of quasiparticles away from the gap-edge and increased recombination rates near the peaks of $\alpha^2F(\omega)$ work to enhance superconductivity, with the notable exception that here $\omega_{\mathrm{pump}} \gg 2\Delta_0$ \cite{eliashberg1972inelastic, chang1977kinetic}. Inclusion of an additional phonon pump term of a similar form to Eq. \eqref{eq:pump} does not have a significant impact on these results. We find, in agreement with experiment, that even for $\omega_{\mathrm{pump}}$ resonant with peaks in $\alpha^2F(\omega)$, a large fluence $F_0$ on the order of $\gtrsim 1\, \mathrm{mJ}/\mathrm{cm}^2$ is required to excite a sufficient quasiparticle density to generate an appreciable enhancement in $\Delta_0$. The time-dependent change in $f(E)$ and $n(\omega)$ is shown in Figure \ref{fig:f_n}, where structure (depletion of quasiparticles and generation of phonons) resonant with the prominent features in $\alpha^2F(\omega)$ appears. For the time when $\Delta_0 > 0$, the peak of the quasiparticle distribution is blue-shifted by $\Delta_0$ due to the opening of the gap. We further find that the photo-induced $\Delta_0$ saturates when the quasiparticle population at the gap-edge becomes smaller than the excited blue-shifted population $f(E=\Delta_0) < f(E=\omega_{\mathrm{pump}}+\Delta_0)$. This effect is illustrated by Figure \ref{fig:f_n}a and discussed in greater detail in the SI \cite{seeSI}. Due to the particularly short-lived superconducting state in K$_3$C$_{60}$, the temporal response will be smeared by the Heisenberg uncertainty principle. 

The enhancement of $\Delta_0$ in our model suggests that conventional electron-phonon-mediated nonequilibrium dynamics can lead to collective excitations, resulting in the potential emergence of a transient superconducting state if time scales are favorable. Additional dynamics not accounted for in this framework could also contribute, including local electronic correlations in K$_3$C$_{60}$ associated with its flat electronic band structure near the Fermi level \cite{K3C60-Arita-2013, doi:10.1126/sciadv.1500568}. Our DFT calculations for the $\alpha^2F(\omega)$ of K$_3$C$_{60}$ underestimates the equilibrium superconducting gap and critical temperature ($T_{\mathrm{c0}} \approx 15\,\mathrm{K} < T_{\mathrm{c0,exp}} \approx20\,\mathrm{K}$), which is likely responsible for the underestimation of the photo-induced gap compared to experiment. Furthermore, future work beyond our particle-hole symmetric treatment and spin-independent occupation in the normal state would provide an even more complete description of the nonequilibrium dynamics. Nevertheless, our calculations clearly demonstrate a mechanism by which conventional electron-phonon-mediated superconductivity can be enhanced through intense, ultrafast mid-infrared excitations resonant with the structure in $\alpha^2F(\omega)$.

The quantitative agreement in both timescale and amplitude with experiments provides strong validation of our theoretical framework. More importantly, it reveals a general mechanism: resonant pump excitation can coherently drive collective pairing interactions through the phonon spectrum, opening pathways to light-enhanced superconductivity across a broad class of conventional superconductors. Materials with prominent high-frequency phonon structure are particularly promising candidates. We identify CaC$_6$ as a promising test case for experimental verification, predicting a photo-response comparable to that of K$_3$C$_{60}$. Together, these results establish a first-principles framework not only as a quantitative explanation of existing experiments, but as a predictive design tool for discovering and optimizing materials for their photo-response.

%TC:ignore

\section{Acknowledgments}

The authors are grateful to Matteo Mitrano, Leonid Levitov, and Boris Spivak for useful discussions. The authors also thank Gian Luca Dolso and Dong-min Kim for their careful review of this manuscript. This work was funded in part by the Defense Sciences Office (DSO) of the Defense Advanced Research Projects Agency (DARPA) (HR0011-24-9-0311). AS and EB acknowledge support from the NSF GRFP. RF acknowledges support from the Alan McWhorter fellowship. PNF acknowledges support from the Austrian Science Fund (FWF) under project DOI 10.55776/ESP8588124. EK, PNF MS, and CH acknowledge support from the Enterprise Science Fund of Intellectual Ventures and usage of computational resources of the lCluster of the Graz University of Technology and of the Austrian Scientific Computing (ASC) infrastructure.

\newpage

\onecolumngrid

\section{Supplementary Information}

\subsection{Kinetic model}

As described in the main text, under the quasiparticle approximation, we can describe the transport in a conductor with the set of kinetic equations for the quasiparticle $f(E)$ and phonon $n(\omega)$ distributions outlined in Eqs. (1a) and (1b). In a normal metal, this approximation results in the Prange-Kadanoff equations \cite{prange1964transport}, and for a superconductor where renormalization effects are weak, one obtains the Chang-Scalapino equations \cite{chang1977kinetic}. The latter is also equivalent to the kinetic equation derived by Eliashberg if the phonons are taken to be at equilibrium \cite{eliashberg1972inelastic}. Here, we extend the Chang-Scalapino equations to incorporate the effect of strong-coupling on the interaction probabilities by solving the kinetic equations self-consistently with the Migdal-Eliashberg equations, replacing the BCS coherence factors and quasiparticle density of states with the corresponding terms computed from the spectral function obtained from the nonequilibrium Migdal-Eliashberg equations. This also requires us to assume that the superconducting gap evolves adiabatically relative to the time scale of quasiparticle interactions, which in general is a good approximation outside of the immediate vicinity of $T_{\mathrm{c,0}}$. To keep the computational workflow feasible, we employ two approximations here: (i) We neglect the time derivative of the self-energy, which should now appear on the left-hand side of the Chang-Scalapino equations \cite{prange1964transport, chang1977kinetic}. The neglect of this term is reasonable for materials where the electron-phonon coupling is not too strong, i.e. $2 \int \mathrm{d}\omega \, \alpha^2F(\omega)/\omega \lesssim 1$. (ii) As $N(\varepsilon)$ is relatively flat within the vicinity of the Fermi level $\varepsilon_{\mathrm{F}}$ for the materials under investigation, we work in the constant density of states approximation.

The kinetic equations (Eqs. (1a) and (1b) in the main text) when fully expanded are
\onecolumngrid\begin{subequations}
\begin{equation}
\begin{aligned}
\label{eq:f1}
    \frac{d f(E)}{d t} = & -\frac{2\pi}{\hbar}\int_{0}^{\infty} d\omega \, \alpha^2 F(\omega)\rho(E+\omega) M_1(E,E+\omega)
    \Bigg\{ f(E)[1-f(E+\omega)]n(\omega) - f(E+\omega)[1-f(E)][n(\omega)+1]\Bigg\} \\
    & -\frac{2\pi}{\hbar}\int_{0}^{\infty} d\omega \, \alpha^2 F(\omega)\rho(E-\omega) M_1(E,E-\omega)
    \Bigg\{ f(E)[1-f(E-\omega)][n(\omega)+1] - [1-f(E)]f(E-\omega)n(\omega)\Bigg\} \\
    & -\frac{2\pi}{\hbar}\int_{0}^{\infty} d\omega \, \alpha^2 F(\omega)\rho(\omega-E)M_2(E,E-\omega)
    \Bigg\{ f(E)f(\omega-E)[n(\omega)+1] - [1-f(E)][1-f(\omega-E)]n(\omega)\Bigg\}
\end{aligned}
\end{equation}
and 
\begin{equation}
\begin{aligned}
\label{eq:n1}
    \frac{d n(\omega)}{d t} = & -\frac{8\pi}{\hbar}\frac{N(\varepsilon_{\mathrm{F}})}{N}\int_{0}^{\infty} d\!E \int_{0}^{\infty} d\!E' \, \alpha^2(\omega)\rho(E)\rho(E') \\
    & \times \Bigg[ M_1(E,E+\omega)\Big\{ f(E)[1-f(E')]n(\omega)  - f(E')[1-f(E)][n(\omega)+1]\Big\}\delta(E+\omega-E') \\
    & \qquad + \tfrac{1}{2}M_2(E,E-\omega)\Big\{ [1-f(E)][1-f(E')]n(\omega) - f(E)f(E')[n(\omega)+1]\Big\}\delta(E+E'-\omega) \Bigg] \\
    & -\frac{n(\omega)-n^{\mathrm{eq}}(\omega)}{\tau_{\mathrm{esc}}},
\end{aligned}
\end{equation}
\end{subequations} 
where $M_1(\varepsilon, \varepsilon')/2 = |u(\tilde{E}(\varepsilon))u(\tilde{E}(\varepsilon')) - v(\tilde{E}(\varepsilon))v(\tilde{E}(\varepsilon'))|^2$ and $M_2(\varepsilon, \varepsilon')/2 = |v(\tilde{E}(\varepsilon))u(\tilde{E}(\varepsilon')) + u(\tilde{E}(\varepsilon))u(\tilde{E}(\varepsilon'))|^2$ are the coherence factors, $u(\varepsilon)$ and $v(\varepsilon)$ are the quasiparticle amplitudes, $\tilde{E}(\varepsilon) = Z(\varepsilon)\sqrt{\varepsilon^2 - \Delta^2(\varepsilon)}$ is the renormalized quasiparticle energy, $N(\varepsilon_{\mathrm{F}})$ is the single-spin electronic density of states at the Fermi level, $N$ is the ion density of the material, $\alpha^2F(\omega)$ is the Eliashberg spectral function, and $F(\omega)$ is the phonon density of states. The quasiparticle amplitudes and quasiparticle density of states $\rho(\varepsilon')$ are obtained by integrating the spectral function 
\begin{equation}
A(\varepsilon, \varepsilon', t) = -\frac{1}{\pi} \Im G^R(\varepsilon, \varepsilon', t),
\end{equation}
where 
\begin{equation}
    G^R(\varepsilon, \varepsilon', t) = \frac{\varepsilon' Z(\varepsilon',t) + \varepsilon}{[\varepsilon' Z(\varepsilon',t)]^2 - \varepsilon^2 - \Delta(\varepsilon',t)^2 + i0^+}
\end{equation} is the retarded Green's function. Specifically, 
\begin{equation}
    \rho(\varepsilon') = \int d\varepsilon A(\varepsilon, \varepsilon')=\Re \Bigg[ \frac{\varepsilon'}  {\sqrt{\varepsilon'^2 - \Delta^2(\varepsilon')}} \Bigg]
\end{equation} and 
\begin{equation}
    |v(\varepsilon)|^2 = \langle n(\varepsilon)\rangle = \int d\omega A(\varepsilon, \omega).
\end{equation} 
By conservation of probability $u^2(\varepsilon) = 1 - v^2(\varepsilon)$. Here, $Z(\varepsilon',t)$ and $\Delta(\varepsilon',t)$ are obtained from solving the Migdal-Eliashberg equations on the real-frequency axis with the nonequilibrium distributions $f(E)$ and $n(\omega)$. We assume for the sake of simplicity in the calculation that the imaginary part of the quasiparticle amplitudes is small. In the limit of $\Delta \rightarrow 0$, Eqs. \eqref{eq:f1} and \eqref{eq:n1} reduce to the Prange-Kadanoff equations for a normal metal. The full real-frequency axis Migdal-Eliashberg equations, which are to be solved self-consistently with Eqs. \eqref{eq:f1} and \eqref{eq:n1}, and the methods used to obtain their solutions are detailed in Refs. \cite{FIXME-THIS-PAPER, simon2025abinitiomodelingnonequilibrium, simon2025ab}. 

An additional remark on the nature of phonon escape is worth further discussion as well. In a thin-film, such as in Ref. \cite{PhysRevB.72.024510}, the phenomenological escape term added at the end of Eq. \eqref{eq:n1} should sufficiently describe the dynamics of ballistic phonon propagation and can be estimated from the acoustic mismatch between the film and substrate. However, the term does not necessarily account for the reintroduction of phonons from the substrate, which is possible, particularly if the cooling power of the testing apparatus is not strong enough to sufficiently cool the substrate in the time of interest \cite{PhysRevB.72.024510}. To account for this, we add an exponentially decaying tail to the calculation of $-\Delta \mathcal{T}/\mathcal{T_0}$ with its amplitude set to 0.2 and decay constant $3\,$ns. The same tail is added to each data so as to not impact the comparison between different temperatures. 

In thick films of materials with short penetration depths, such as the study on LaH$_{10}$ \cite{Wu2024-rm}, the observed response is localized to the surface, and hence the phenomenological escape term is also a reasonable approximation, except it should be expected that $\tau_{\mathrm{esc}}$ is significantly smaller since there is no appreciable acoustic mismatch between the excited and equilibrium material. Indeed, this is the case, where the experimental data is well-described by $\tau_{\mathrm{esc}} = 500\,$ps for Pb and for LaH$_{10}$ $\tau_{\mathrm{esc}}$ is set to be a monotonically decreasing function with increasing fluence $F_0$ that varies from $\tau_{\mathrm{esc}}=50\,$fs, $200\,$fs, and $500\,$fs for $F_0=2.6\,$mJ/cm$^2$, $4.5\,$mJ/cm$^2$, $6.2\,$mJ/cm$^2$, respectively. The reintroduction of phonons from non-optically excited region is a stronger effect than for phonons entering from a substrate, where phonons move more freely in the bulk, and hence our calculations are expected to underestimate the value $\Delta R/R_0$ saturates to in LaH$_{10}$. At higher fluences, a much denser, hotter nonequilibrium phonon bath is created, which increases anharmonic phonon–phonon scattering and reabsorption and recycling of high-energy phonons that can break pairs. Both effects reduce the mean free path and keep energy in the pair-breaking window locally for longer, which appears as a longer effective $\tau_{\mathrm{esc}}$. At lower fluences, these additional scattering channels are weaker and transport is closer to ballistic in the sense of longer mean free paths, so the same near-surface region drains faster and $\tau_{\mathrm{esc}}$ decreases. A more complete treatment of the bulk phonon dynamics would remedy this issue, but is outside of the scope of this study. 

\subsection{Optical model}
As described in the text, the pump-probe experiments performed on Pb and LaH$_{10}$ can be modeled using the Fresnel equation for normal incidence, resulting in 
\begin{equation}
\label{eq:t}
    t = \frac{4 N_1 N_2}{(N_1 + N_2)(N_2 + N_3)e^{-i\phi} - (N_2-N_3)(N_2-N_1)e^{i\phi}}
\end{equation}
for the transmission coefficient and 
\begin{equation}
\label{eq:r}
r = \frac{(N_1 - N_2)(N_2 + N_3)e^{-i\phi} + (N_2-N_3)(N_1+N_2)e^{i\phi}}{(N_1+N_2)(N_2+N_3)e^{-i\phi} - (N_2-N_3)(N_2-N_1)e^{i\phi}}
\end{equation}
for the reflection coefficient. The reflectance and transmittance are then determined from
\begin{equation}
    T = \frac{N_3}{N_1}|t|^2,~\mathrm{and}~R=|r|^2.
\end{equation} Here, $N_1(\omega),~N_2(\omega, t),$ and $N_3(\omega)$ are the indices of refraction of the first, second, and third layer, with their frequency and time dependencies suppressed in Eqs. \eqref{eq:t} and \eqref{eq:r} for clarity and
\begin{equation}
    \phi = N_2 \frac{\omega d}{c}.
\end{equation} 
The values for $N_1(\omega)$ and $N_3(\omega)$ are taken from the experimental data, whereas the time-dependent index of the sample $N_2(\omega,t)$ is computed from the Kubo formula as described in the main text. The penetration depth at frequency $\omega$ can be estimated from
\begin{equation}
    \delta(\omega) = \frac{c}{2\omega \, \Im n(\omega)}, 
\end{equation} which we confirm matches the previously reported penetration depths for the samples under study. 

The observed optical response will also be convolved with the pulse shape of the probe. In Refs. \cite{PhysRevB.72.024510, Wu2024-rm}, $\tau_{\mathrm{pr}} = 200\,$ps for Pb and $\tau_{\mathrm{pr}} = 70\,$fs for LaH$_{10}$; however, for LaH$_{10}$ we find improved agreement with the rising edge of the experimental data for a slightly larger time constant of $250\,$fs. % We provide in Figure \ref{fig:rawPb-LaH10} the response of the corresponding response of Pb and LaH$_{10}$ before convolution is performed. 

\subsection{Computational Details}
K$_3$C$_{60}$: Electron-phonon calculations were performed using the all-electron FHI-aims code package \cite{Blum2009-fhiaims, Box2023-fhiaims-fric} and the Perdew-Burke-Ernzerhof (PBE) exchange-correlation functional \cite{perdew_generalized_PBE_1996}. 
Basis functions and element-specific numerical parameters were taken from the 2020 lightdense default preset of FHI-aims. The Brillouin zone was sampled using a $\mathbf{k}$-grid density of 6\,\AA$^{-1}$. Electronic occupations were treated using Methfessel–Paxton smearing~\cite{Methfessel_PRB_1989_smearing} of 0.3\,eV. In the electron-phonon coupling calculation, a friction broadening width of 0.3\,eV was employed. Calculations were carried out on a unit cell comprising 63 atoms and did not account for the orientational disorder of the C$_{60}$ units observed in experiment \cite{Stephens1991}.
This simplification may account for the four phonon modes with imaginary frequencies obtained in our calculations. The occurrence of imaginary phonon modes is consistent with findings from other ab-initio studies~\cite{K3C60-Arita-2013}. A more detailed investigation of the crystal structure would require larger supercells, which is beyond the scope of this work.

Pb: See Ref. \cite{simon2025abinitiomodelingnonequilibrium}.

LaH$_{10}$: DFT and DFPT calculations were performed using the Quantum Espresso package \cite{QE-2017, QE-2020}. Exchange–correlation effects were described using the PBE functional \cite{perdew_generalized_PBE_1996}, together with scalar-relativistic optimized norm-conserving Vanderbilt pseudopotentials \cite{vanbilt_pseudo_hamann_2013_PhysRevB.88.085117}. A $12\times12\times12$ $\mathbf{k}$-grid, an energy cutoff of 90\,Ry, and a Methfessel-Paxton smearing \cite{Methfessel_PRB_1989_smearing} of 0.01\,Ry were applied. 
To account for anharmonic and quantum ionic effects, which are necessary to stabilize LaH$_{10}$ at 165\,GPa in the $Fm\bar{3}m$ phase \cite{Errea2020-SSCHA-LaH10}, the stochastic self-consistent harmonic approximation (SSCHA) \cite{Bianco2017-SSCHA1, Monacelli2018-SSCHA2, Monacelli2021-SSCHA3} was applied.
SSCHA calculations were accelerated by a moment tensor potential (MTP), generated using the MLIP package \cite{Shapeev2016-MTP1, Novikov2021-MTP2}. Our workflow closely follows that described in Ref.~\cite{Lucrezi2023}. The best performance on the validation set was obtained with an MTP of level 26, trained on DFT energies, forces, and stresses from 75 and 25 supercell configurations of sites $2 \times 2 \times 2$ and $3 \times 3 \times 3$ respectively, and a maximum number of 500 iterations in the optimization algorithm. Training and validation sets were randomly chosen out of the initial generations of the SSCHA run, which were performed using DFT. The electron-phonon coupling was recomputed based on the dynamical matrices of a $3\times3\times3$ supercell, which included 50000 configurations in the SSCHA sampling.

CaC$_{6}$: We carried out DFT calculations with the Quantum \textsc{ESPRESSO} suite~\cite{QE-2017, QE-2020}. Scalar-relativistic optimized norm-conserving Vanderbilt (ONCV) pseudopotentials~\cite{ONCV1,ONCV2} were adopted from the \textsc{PseudoDojo} library~\cite{setten2018}. The exchange-correlation energy was described within PBE functional~\cite{perdew_generalized_PBE_1996}. Self-consistent-field calculations were converged to 10$^{-10}$\,Ry, employing a plane-wave kinetic-energy cutoff of 100\,Ry. Brillouin-zone integrations were performed on a $\Gamma$-centered Monkhorst-Pack \textbf{k}-point mesh~\cite{monkhorst1976} of 18$\times$18$\times$18, together with Methfessel-Paxton smearing~\cite{Methfessel_PRB_1989_smearing} of 0.04\,Ry. With this setup, the total energies are converged to better than 1\,meV/atom. Both lattice constants and internal atomic coordinates were relaxed until the total-energy variation and residual forces were below 10$^{-7}$\,Ry and 10$^{-6}$\,Ry/a$_0$, respectively. Electron-phonon matrix elements were evaluated on uniform 6$\times$6$\times$6 phonon grids. For the phonon calculations, the self-consistency threshold was set to 10$^{-16}$\,Ry. The electron-phonon matrix elements were further integrated over a denser 36$\times$36$\times$36 \textbf{k}-point grid, using 50 double-delta smearing parameters between 0.001 and 0.05\,Ry. The values discussed in the main text were obtained with a smearing of 0.05\,Ry and a phonon smearing of 0.30\,THz for the \textbf{q}-point integration.

\section{Equilibrium gap}

\begin{figure}[H]
    \centering
    \includegraphics[width=0.6\linewidth]{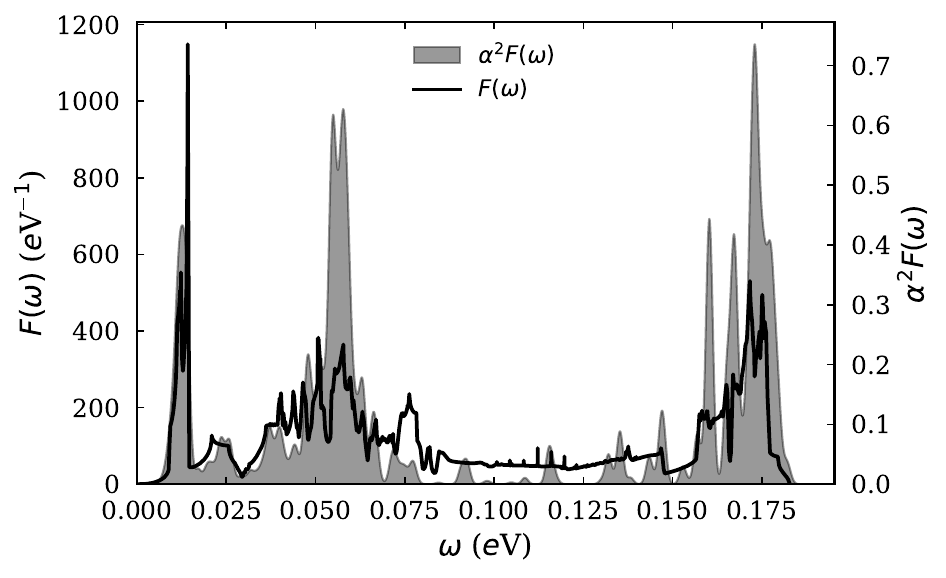}
    \caption{The Eliashberg spectral function $\alpha^2F(\omega)$ and phonon density of states $F(\omega)$ for CaC$_6$.}
    \label{fig:CaC6-DFT}
\end{figure}

\begin{figure}[htbp]
    \centering
    \begin{subfigure}[t]{0.45\textwidth}
        \centering
        \includegraphics[width=\textwidth]{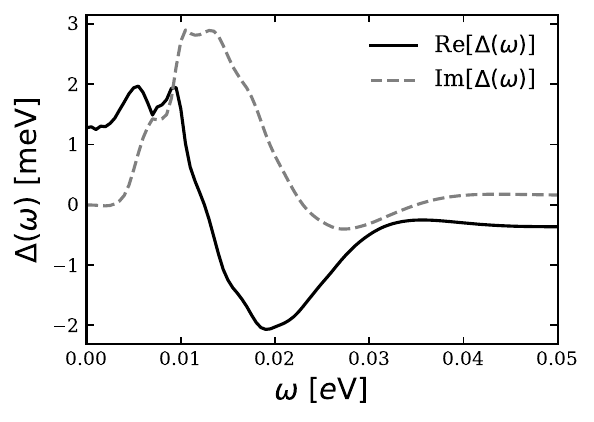}
        \subcaption{The equilibrium superconducting gap $\Delta(\omega)$ for Pb at $T=3.8\,\mathrm{K}$.}
        \label{fig:Pb_gap}
    \end{subfigure}
    \hfill
    \begin{subfigure}[t]{0.45\textwidth}
        \centering
        \includegraphics[width=\textwidth]{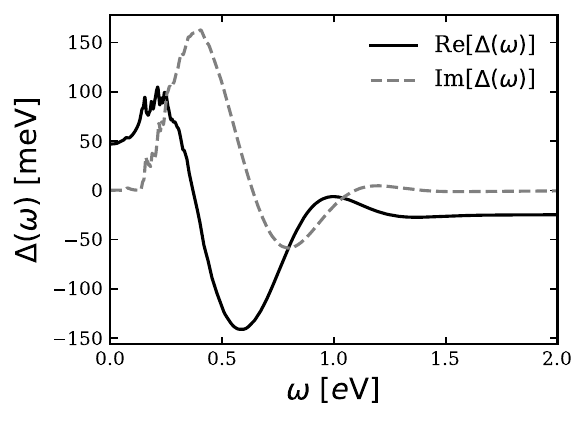}
        \subcaption{The equilibrium superconducting gap $\Delta(\omega)$ for LaH$_{10}$ at $T=77\,\mathrm{K}$.}
        \label{fig:LaH10_gap}
    \end{subfigure}
    \hfill
    \begin{subfigure}[t]{0.45\textwidth}
        \centering
        \includegraphics[width=\textwidth]{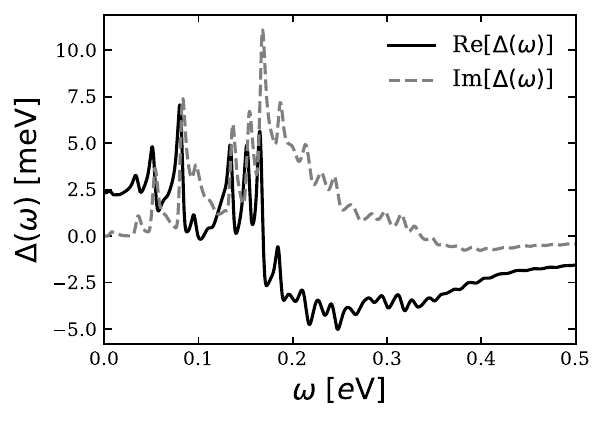}
        \subcaption{The equilibrium superconducting gap $\Delta(\omega)$ for K$_3$C$_{60}$ at $T=1\,\mathrm{K}$.}
        \label{fig:K3C60_gap}
    \end{subfigure}
    \hfill
    \begin{subfigure}[t]{0.45\textwidth}
        \centering
        \includegraphics[width=\textwidth]{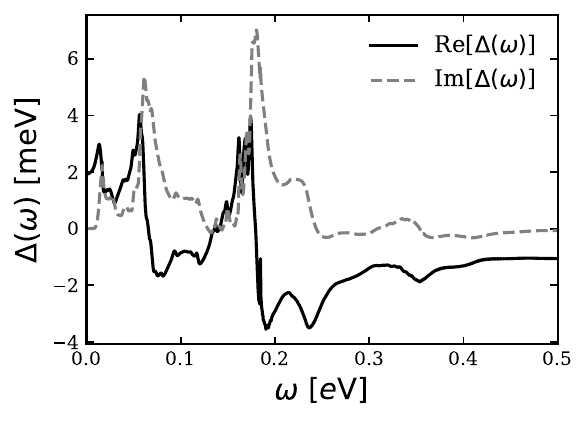}
        \subcaption{The equilibrium superconducting gap $\Delta(\omega)$ for CaC$_6$ at $T=1\,\mathrm{K}$.}
        \label{fig:CaC6_gap}
    \end{subfigure}
    % \vspace{-15pt}
    \caption{Equilibrium superconducting gap $\Delta(\omega)$ for the materials under study.}
    \label{fig:all}
\end{figure}

In Figure \ref{fig:CaC6-DFT}, we show the DFT computed $\alpha^2F(\omega)$ and $F(\omega)$ for CaC$_6$. In Figures \ref{fig:Pb_gap}, \ref{fig:LaH10_gap}, \ref{fig:K3C60_gap}, and \ref{fig:CaC6_gap}, we show the real-frequency equilibrium gap of Pb, LaH$_{10}$, K$_3$C$_{60}$, and CaC$_6$, respectively, computed with the real-frequency-axis Migdal-Eliashberg solver \cite{FIXME-THIS-PAPER}.

\section{Variation of the penetration depth in nonequilibrium}

\begin{figure}[htbp]
    \centering
    \begin{subfigure}[t]{0.45\textwidth}
        \centering
        \includegraphics[width=\textwidth]{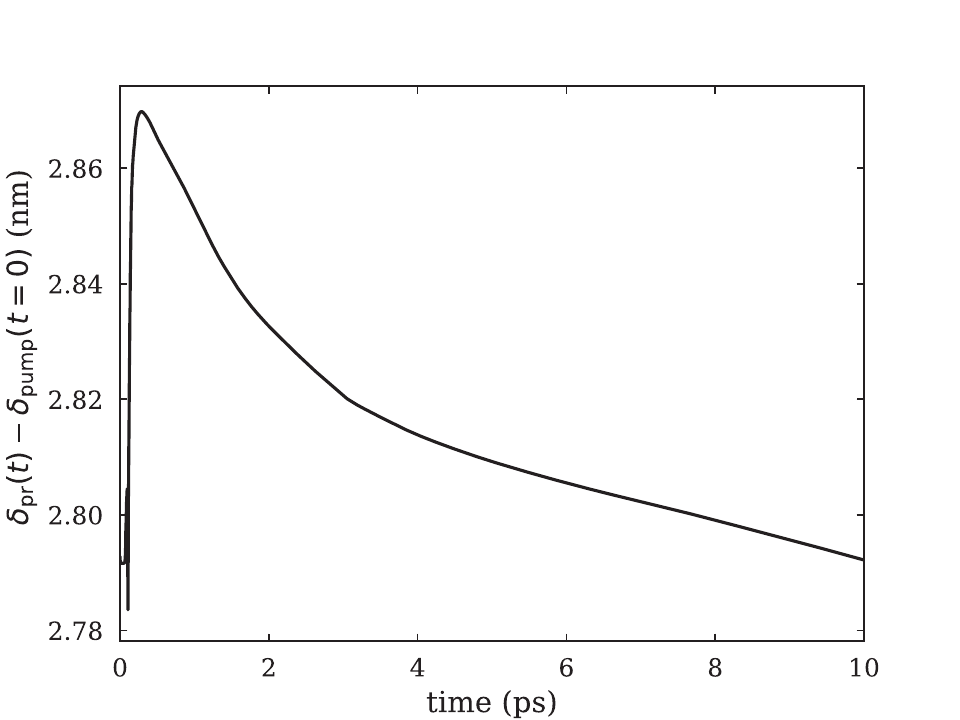}
        \caption{The variation of the probe penetration depth $\delta_{\mathrm{pr}}(t)$ from the equilibrium pump penetration depth $\delta_{\mathrm{pump}}(t=0)$ after LaH$_{10}$ is pumped at $t=0$.}
        \label{fig:penetration_depths}
    \end{subfigure}
    \hfill
    \begin{subfigure}[t]{0.45\textwidth}
        \centering
        \includegraphics[width=\textwidth]{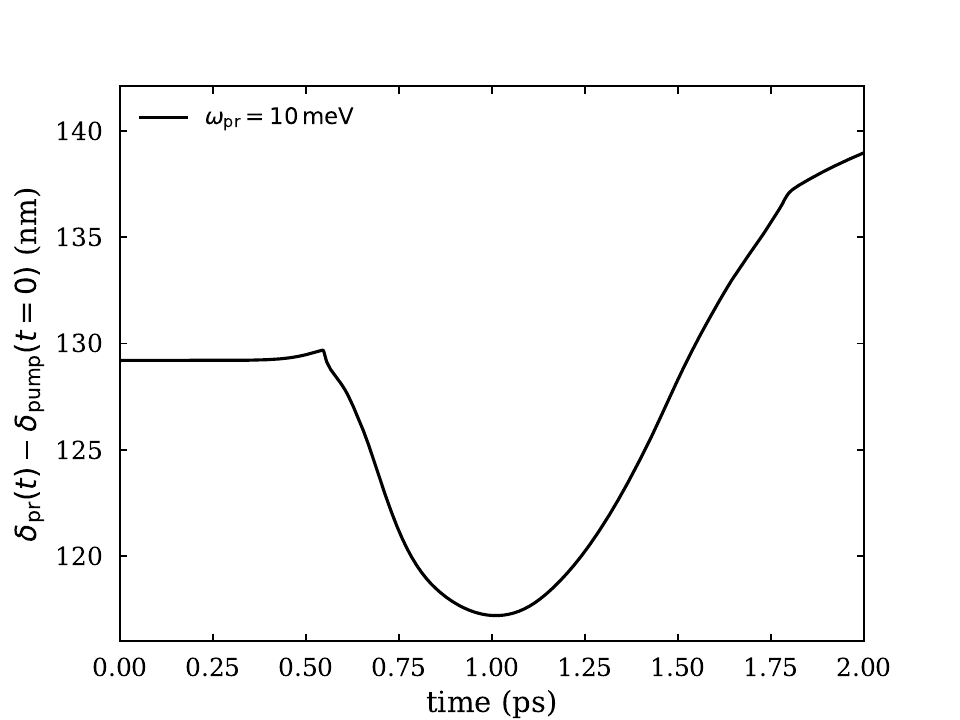}
        \caption{The variation of the probe penetration depth $\delta_{\mathrm{pr}}(t)$ from the equilibrium pump penetration depth $\delta_{\mathrm{pump}}(t=0)$ after K$_3$C$_{60}$ is pumped at $t=1\,$ps.}
        \label{fig:K3C60_penetration_depth}
    \end{subfigure}
    % \vspace{-15pt}
    \caption{Variation of the probe penetration depth $\delta_{\mathrm{pr}}(\omega,t)$ in LaH$_{10}$ and K$_3$C$_{60}$.}
    \label{fig:pen}
\end{figure}

The time-dependent penetration depth can be estimated by solving for the absorption coefficient as a function of the time-dependent index of refraction $n(\omega,t)$. If the penetration depth of the pump and probe differ, this can significantly modify the observed optical response of a material \cite{PhysRevLett.130.146002}. Thus, the most general solution should account for this by also considering the spatial dependence of $n(\omega,t)$. In the main text, we calculated $-\Delta \mathcal{T}/\mathcal{T}_0$ and $\Delta R/R_0$ for Pb and LaH$_{10}$, respectively. The Pb film is taken to be in the thin-film limit as is consistent with the experimental setup in Ref. \cite{PhysRevB.72.024510}, and hence any excitation can be taken to be uniform across the thickness, removing the ambiguity associated with different penetration depths. For LaH$_{10}$, we must compute the difference in the penetration depth for the experimental setup in Ref. \cite{Wu2024-rm}, where $\omega_{\mathrm{pump}} = 2\,e\mathrm{V}$ and $\omega_{\mathrm{pr}} = 1.5\,e\mathrm{V}$. Due to the similar frequencies of the pump and probe, the difference in the penetration depths at equilibrium is only approximately 3$\,$nm, with the pump penetration depth $\delta_{\mathrm{pump}}$ approximately equal to 20$\,$nm. Upon excitation by the pump at the fluence of $F_0=6.2\,$mJ/cm$^2$, this only varies by at most a few angstroms as shown in Figure \ref{fig:penetration_depths}. From this, we may conclude that in LaH$_{10}$, it is a relatively robust approximation to assume the probe will see primarily the excited material and can neglect the spatial dependence of $n(\omega,t)$. To illustrate a case where this approximation is not adequate, we also include in Figure \ref{fig:K3C60_penetration_depth} the time-dependent variation in the probe penetration depth for K$_3$C$_{60}$ at a probe frequency $\omega_{\mathrm{pr}} = 10\,\mathrm{m}e\mathrm{V}$ that lies near the superconducting gap. For K$_3$C$_{60}$, the difference between the probe penetration depth and the equilibrium pump penetration depth can vary on the order of 100 nanometers and varies by an additional several tens of nanometers in the nonequilibrium state. For reference, $\delta_{\mathrm{pump}}$ at $\omega_{\mathrm{pump}}=170\,\mathrm{m}e\mathrm{V}$ is approximately 200$\,$nm for K$_3$C$_{60}$.

\section{Fluence and pump frequency dependence of the photo-induced superconductivity in K$_3$C$_{60}$}

\begin{figure}[H]
    \centering
    \includegraphics[width=1\linewidth]{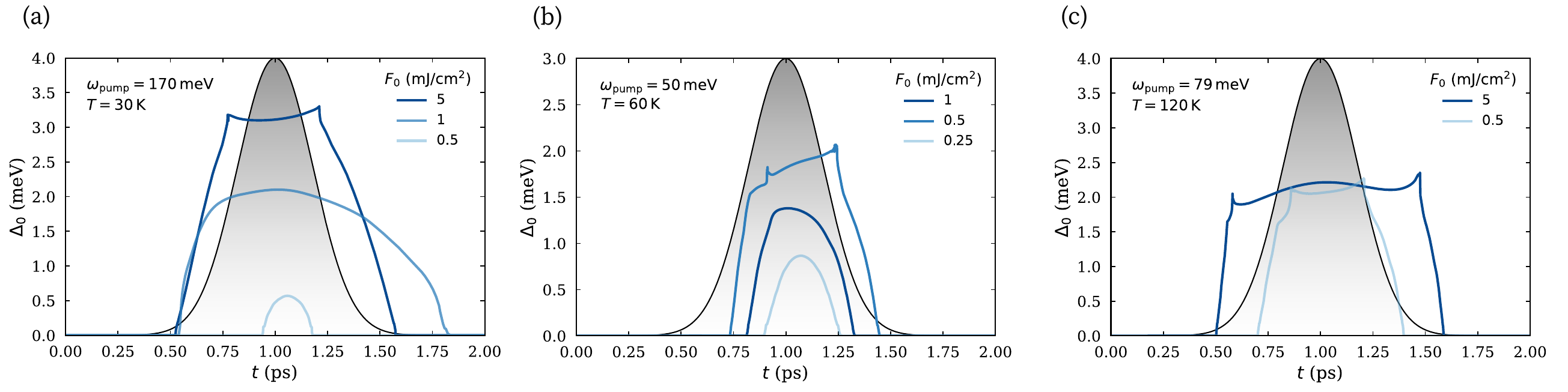}
    \caption{The fluence $F_0$ dependence of the photo-induced gap $\Delta_0$ at $T=30\,\mathrm{K}$ and $\omega_{\mathrm{pump}}=170\,\mathrm{m}e\mathrm{V}$, $T=120\,\mathrm{K}$ and $\omega_{\mathrm{pump}}=79\,\mathrm{m}e\mathrm{V}$, and $T=60\,\mathrm{K}$ and $\omega_{\mathrm{pump}}=50\,\mathrm{m}e\mathrm{V}$.}
    \label{fig:K3C60-fluence}
\end{figure}

The calculated $F_0$ dependence of the photo-induced state in K$_3$C$_{60}$ at $\omega_{\mathrm{pump}} = 170\,$m$e$V and $T=30\,$K is shown in Figure \ref{fig:K3C60-fluence}. Upon reaching $5\,$mJ/cm$^2$, the photo-induced gap saturates to a maximum value of $2\Delta_0\approx7\,$m$e$V. We discuss the origin of this saturation and the sharp gradients in $\Delta_0$ in the following section. Interestingly, for $\omega_{\mathrm{pump}}=170\,$m$e$V, the relaxation of the saturated gap is faster than for the lower fluence $1\,$mJ/cm$^2$, indicating that the generation of excess quasiparticles is detrimental to the stability of the photo-induced state. The actual fluence at which this occurs is necessarily larger, since we neglect quasiparticle and phonon diffusion in this model, which would significantly reduce the energy density on $>$ps time scales.  

In addition to the results shown in the main text for $\omega_{\mathrm{pump}} = 170\,\mathrm{m}e\mathrm{V}$, we also compute the response to $\omega_{\mathrm{pump}} = 50\,\mathrm{m}e\mathrm{V}$ and $\omega_{\mathrm{pump}} = 79\,\mathrm{m}e\mathrm{V}$ at higher temperatures. Evidently, at $\omega_{\mathrm{pump}}$ resonant with lower frequency resonances in $\alpha^2F(\omega)$ a smaller $F_0$ is required to reach saturation of $\Delta_0$, which is in agreement with recent experimental data \cite{Rowe2023-ab}. The shape of saturated $\Delta_0$ also clearly has a dependence on $\omega_{\mathrm{pump}}$.

\begin{figure}[H]
    \centering
    \includegraphics[width=0.8\linewidth]{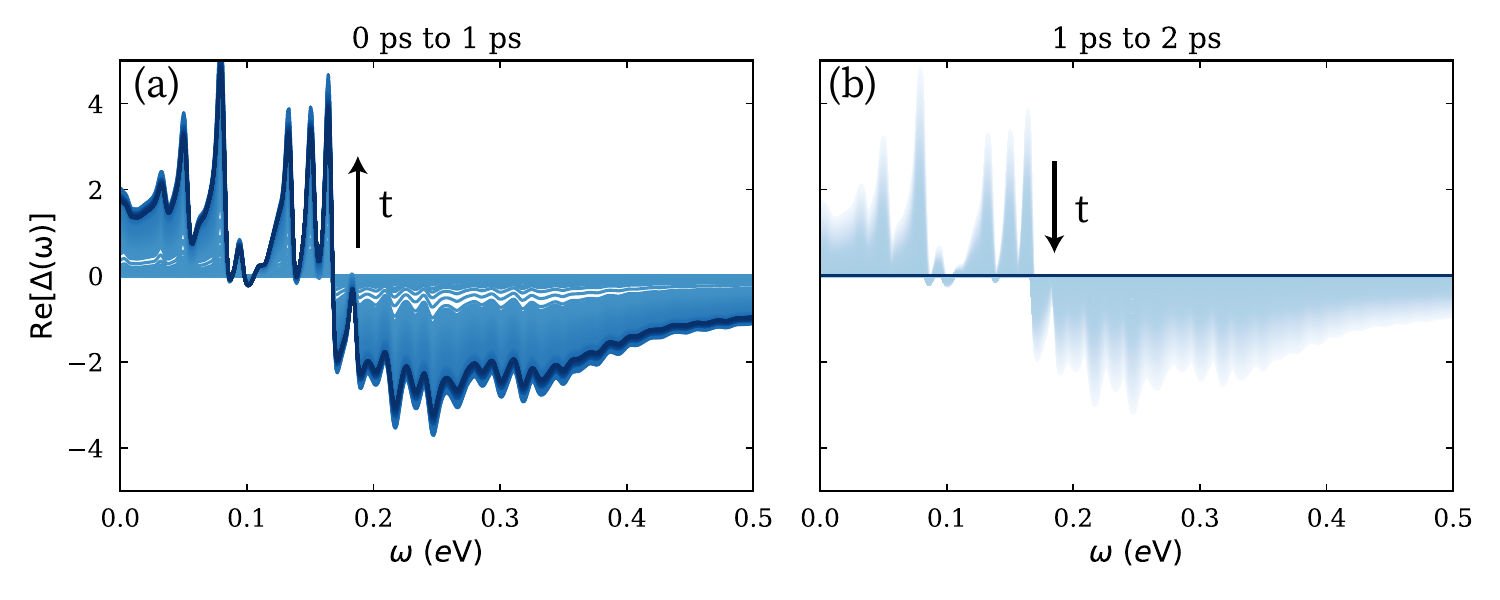}
    \caption{The frequency dependence of the real part of the time-dependent superconducting gap $\Delta(\omega,t)$ from (a) $t=0\,$ps to $t=1\,$ps and (b) $t=1\,$ps to $t=2\,$ps. The color progression from light to dark and arrows indicate the direction of the evolution of the magnitude of the gap as time progresses.}
    \label{fig:eqGaps}
\end{figure}

In Figure \ref{fig:eqGaps}, we display the evolution of $\Delta(\omega,t)$ for K$_3$C$_{60}$.

\section{Dynamics of the excited quasiparticle population in the presence of the photo-induced gap and the origin of gap saturation}
\label{sec:f_n}

Upon formation of the photo-induced gap, the quasiparticle peak is blue shifted to $\omega_{\mathrm{pump}} + \Delta_0$. To clearly depict this, we provide supplemental videos illustrating the time evolution of $\Delta_0$, $f(\omega)$, and $n(\omega)$. The saturation of $\Delta_0$ then occurs when the low-frequency quasiparticles are depleted. That is, $f(\omega=\Delta_0(t)) \ll f(\omega=\omega_{\mathrm{pump}}+\Delta_0(t))$. Sharp gradients appear as $\Delta_0(t)$ traverses strongly depleted regions of $f(\omega)$, which also results in a non-trivial dependence of $\Delta_0$ on $\omega_{\mathrm{pump}}$ and $\alpha^2F(\omega)$. However, these sharp gradients should not be expected to be physical, as these features should be washed out over time scales of $\hbar/\Delta_0$. 

\begin{figure}[H]
    \centering
    \includegraphics[width=0.5\linewidth]{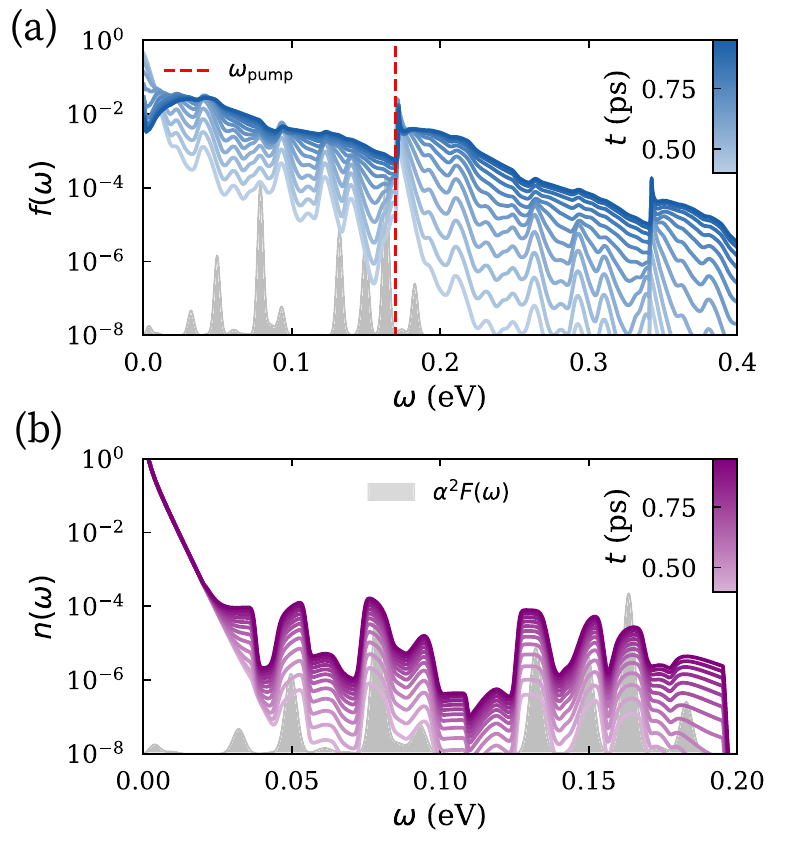}
    \caption{The (a) quasiparticle distribution function $f(\omega,t)$ and (b) phonon distribution function from $t=0.4\,$ps to $1\,$ps for K$_3$C$_{60}$ when pumped with $\omega_{\mathrm{pump}}=170\,$m$e$V. A clear discontinuity in $f(\omega)$ is present at integer multiples of $\omega_{\mathrm{pump}}.$}
    \label{fig:K3C60_f_n}
\end{figure}

\begin{figure}[H]
    \centering
    \includegraphics[width=0.5\linewidth]{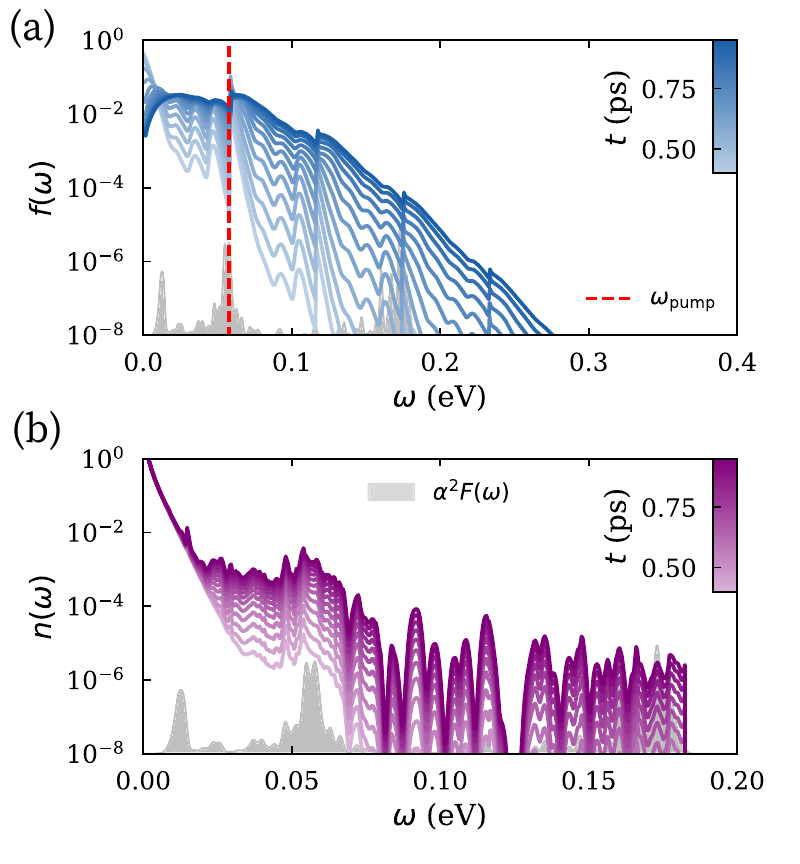}
    \caption{The (a) quasiparticle distribution function $f(\omega,t)$ and (b) phonon distribution function from $t=0.4\,$ps to $1\,$ps for CaC$_{6}$ when pumped with $\omega_{\mathrm{pump}}=58\,$m$e$V.}
    \label{fig:CaC6_f_n}
\end{figure}

In Figures \ref{fig:K3C60_f_n} and \ref{fig:CaC6_f_n}, the formed structure is clearly resonant with $\alpha^2F(\omega)$, with a sharp discontinuity appearing at $\omega_{\mathrm{pump}}$ and its higher harmonics. On the other hand, the shape of $n(\omega)$ is nearly entirely dominated by $\alpha^2F(\omega)$ and $F(\omega)$.

\section{Material parameters}

\begin{table}[H]
    \centering
    \renewcommand{\arraystretch}{1.2}
    \begin{tabular}{l c c c c}
        \toprule
        \textbf{Symbol} & \textbf{Pb} & \textbf{LaH$_{10}$} & \textbf{K$_3$C$_{60}$} & \textbf{CaC$_6$} \\ \midrule
        $N~[\rm{nm}^{-3}]$ & 33 & 350 & 87 & 95 \\
        $\rho_N~[\Omega\cdot\rm{nm}]$ & 350 & 3000 & 20000 & 460 \\
        $N(0)~[e \rm V^{-1}nm^{-3}]$ & 5.895 & 10 & 5.2 & 20.5 \\
        % $\tau_{\rm{esc}}~[\rm{ps}]$ & 100 & 1 & 100 & 100 \\
        $T_{\mathrm{c,0}}~[\rm{K}]$ & 8 & 250 & 15 & 11 \\ \bottomrule
    \end{tabular}
    \caption{Material parameters used in simulations. The equilibrium critical temperature $T_{\mathrm{c,0}}$ was determined from the DFT calculations and real-frequency-axis Migdal-Eliashberg solver and are generally in agreement with experiment, with the exception of the underestimation of $T_{\mathrm{c,0}}$ in K$_3$C$_{60}$ by $5\,$K \cite{FIXME-THIS-PAPER}.}
    \label{tab:material_parameters}
\end{table}

%TC:endignore

% \nocite{*}



\begin{thebibliography}{68}%
\makeatletter
\providecommand \@ifxundefined [1]{%
 \@ifx{#1\undefined}
}%
\providecommand \@ifnum [1]{%
 \ifnum #1\expandafter \@firstoftwo
 \else \expandafter \@secondoftwo
 \fi
}%
\providecommand \@ifx [1]{%
 \ifx #1\expandafter \@firstoftwo
 \else \expandafter \@secondoftwo
 \fi
}%
\providecommand \natexlab [1]{#1}%
\providecommand \enquote  [1]{``#1''}%
\providecommand \bibnamefont  [1]{#1}%
\providecommand \bibfnamefont [1]{#1}%
\providecommand \citenamefont [1]{#1}%
\providecommand \href@noop [0]{\@secondoftwo}%
\providecommand \href [0]{\begingroup \@sanitize@url \@href}%
\providecommand \@href[1]{\@@startlink{#1}\@@href}%
\providecommand \@@href[1]{\endgroup#1\@@endlink}%
\providecommand \@sanitize@url [0]{\catcode `\\12\catcode `\$12\catcode `\&12\catcode `\#12\catcode `\^12\catcode `\_12\catcode `\%12\relax}%
\providecommand \@@startlink[1]{}%
\providecommand \@@endlink[0]{}%
\providecommand \url  [0]{\begingroup\@sanitize@url \@url }%
\providecommand \@url [1]{\endgroup\@href {#1}{\urlprefix }}%
\providecommand \urlprefix  [0]{URL }%
\providecommand \Eprint [0]{\href }%
\providecommand \doibase [0]{https://doi.org/}%
\providecommand \selectlanguage [0]{\@gobble}%
\providecommand \bibinfo  [0]{\@secondoftwo}%
\providecommand \bibfield  [0]{\@secondoftwo}%
\providecommand \translation [1]{[#1]}%
\providecommand \BibitemOpen [0]{}%
\providecommand \bibitemStop [0]{}%
\providecommand \bibitemNoStop [0]{.\EOS\space}%
\providecommand \EOS [0]{\spacefactor3000\relax}%
\providecommand \BibitemShut  [1]{\csname bibitem#1\endcsname}%
\let\auto@bib@innerbib\@empty
%</preamble>
\bibitem [{\citenamefont {Basov}\ \emph {et~al.}(2017)\citenamefont {Basov}, \citenamefont {Averitt},\ and\ \citenamefont {Hsieh}}]{basov2017towards}%
  \BibitemOpen
  \bibfield  {author} {\bibinfo {author} {\bibfnamefont {D.}~\bibnamefont {Basov}}, \bibinfo {author} {\bibfnamefont {R.}~\bibnamefont {Averitt}},\ and\ \bibinfo {author} {\bibfnamefont {D.}~\bibnamefont {Hsieh}},\ }\bibfield  {title} {\bibinfo {title} {Towards properties on demand in quantum materials},\ }\href@noop {} {\bibfield  {journal} {\bibinfo  {journal} {Nature materials}\ }\textbf {\bibinfo {volume} {16}},\ \bibinfo {pages} {1077} (\bibinfo {year} {2017})}\BibitemShut {NoStop}%
\bibitem [{\citenamefont {De~La~Torre}\ \emph {et~al.}(2021)\citenamefont {De~La~Torre}, \citenamefont {Kennes}, \citenamefont {Claassen}, \citenamefont {Gerber}, \citenamefont {McIver},\ and\ \citenamefont {Sentef}}]{de2021colloquium}%
  \BibitemOpen
  \bibfield  {author} {\bibinfo {author} {\bibfnamefont {A.}~\bibnamefont {De~La~Torre}}, \bibinfo {author} {\bibfnamefont {D.~M.}\ \bibnamefont {Kennes}}, \bibinfo {author} {\bibfnamefont {M.}~\bibnamefont {Claassen}}, \bibinfo {author} {\bibfnamefont {S.}~\bibnamefont {Gerber}}, \bibinfo {author} {\bibfnamefont {J.~W.}\ \bibnamefont {McIver}},\ and\ \bibinfo {author} {\bibfnamefont {M.~A.}\ \bibnamefont {Sentef}},\ }\bibfield  {title} {\bibinfo {title} {Colloquium: Nonthermal pathways to ultrafast control in quantum materials},\ }\href@noop {} {\bibfield  {journal} {\bibinfo  {journal} {Reviews of Modern Physics}\ }\textbf {\bibinfo {volume} {93}},\ \bibinfo {pages} {041002} (\bibinfo {year} {2021})}\BibitemShut {NoStop}%
\bibitem [{\citenamefont {Oka}\ and\ \citenamefont {Kitamura}(2019)}]{floquet}%
  \BibitemOpen
  \bibfield  {author} {\bibinfo {author} {\bibfnamefont {T.}~\bibnamefont {Oka}}\ and\ \bibinfo {author} {\bibfnamefont {S.}~\bibnamefont {Kitamura}},\ }\bibfield  {title} {\bibinfo {title} {Floquet engineering of quantum materials},\ }\href {https://doi.org/https://doi.org/10.1146/annurev-conmatphys-031218-013423} {\bibfield  {journal} {\bibinfo  {journal} {Annual Review of Condensed Matter Physics}\ }\textbf {\bibinfo {volume} {10}},\ \bibinfo {pages} {387} (\bibinfo {year} {2019})}\BibitemShut {NoStop}%
\bibitem [{\citenamefont {Kauranen}\ and\ \citenamefont {Zayats}(2012)}]{Kauranen2012-qr}%
  \BibitemOpen
  \bibfield  {author} {\bibinfo {author} {\bibfnamefont {M.}~\bibnamefont {Kauranen}}\ and\ \bibinfo {author} {\bibfnamefont {A.~V.}\ \bibnamefont {Zayats}},\ }\bibfield  {title} {\bibinfo {title} {Nonlinear plasmonics},\ }\href@noop {} {\bibfield  {journal} {\bibinfo  {journal} {Nat. Photonics}\ }\textbf {\bibinfo {volume} {6}},\ \bibinfo {pages} {737} (\bibinfo {year} {2012})}\BibitemShut {NoStop}%
\bibitem [{\citenamefont {Mitrano}\ \emph {et~al.}(2016)\citenamefont {Mitrano}, \citenamefont {Cantaluppi}, \citenamefont {Nicoletti}, \citenamefont {Kaiser}, \citenamefont {Perucchi}, \citenamefont {Lupi}, \citenamefont {Di~Pietro}, \citenamefont {Pontiroli}, \citenamefont {Ricc{\`o}}, \citenamefont {Clark}, \citenamefont {Jaksch},\ and\ \citenamefont {Cavalleri}}]{Mitrano2016-pk}%
  \BibitemOpen
  \bibfield  {author} {\bibinfo {author} {\bibfnamefont {M.}~\bibnamefont {Mitrano}}, \bibinfo {author} {\bibfnamefont {A.}~\bibnamefont {Cantaluppi}}, \bibinfo {author} {\bibfnamefont {D.}~\bibnamefont {Nicoletti}}, \bibinfo {author} {\bibfnamefont {S.}~\bibnamefont {Kaiser}}, \bibinfo {author} {\bibfnamefont {A.}~\bibnamefont {Perucchi}}, \bibinfo {author} {\bibfnamefont {S.}~\bibnamefont {Lupi}}, \bibinfo {author} {\bibfnamefont {P.}~\bibnamefont {Di~Pietro}}, \bibinfo {author} {\bibfnamefont {D.}~\bibnamefont {Pontiroli}}, \bibinfo {author} {\bibfnamefont {M.}~\bibnamefont {Ricc{\`o}}}, \bibinfo {author} {\bibfnamefont {S.~R.}\ \bibnamefont {Clark}}, \bibinfo {author} {\bibfnamefont {D.}~\bibnamefont {Jaksch}},\ and\ \bibinfo {author} {\bibfnamefont {A.}~\bibnamefont {Cavalleri}},\ }\bibfield  {title} {\bibinfo {title} {Possible light-induced superconductivity in {K3C60} at high temperature},\ }\href@noop {} {\bibfield  {journal} {\bibinfo  {journal} {Nature}\ }\textbf {\bibinfo {volume} {530}},\ \bibinfo
  {pages} {461} (\bibinfo {year} {2016})}\BibitemShut {NoStop}%
\bibitem [{\citenamefont {Guo}\ \emph {et~al.}(2024)\citenamefont {Guo}, \citenamefont {Driver}, \citenamefont {Beauvarlet}, \citenamefont {Cesar}, \citenamefont {Duris}, \citenamefont {Franz}, \citenamefont {Alexander}, \citenamefont {Bohler}, \citenamefont {Bostedt}, \citenamefont {Averbukh}, \citenamefont {Cheng}, \citenamefont {DiMauro}, \citenamefont {Doumy}, \citenamefont {Forbes}, \citenamefont {Gessner}, \citenamefont {Glownia}, \citenamefont {Isele}, \citenamefont {Kamalov}, \citenamefont {Larsen}, \citenamefont {Li}, \citenamefont {Li}, \citenamefont {Lin}, \citenamefont {McCracken}, \citenamefont {Obaid}, \citenamefont {O'Neal}, \citenamefont {Robles}, \citenamefont {Rolles}, \citenamefont {Ruberti}, \citenamefont {Rudenko}, \citenamefont {Slaughter}, \citenamefont {Sudar}, \citenamefont {Thierstein}, \citenamefont {Tuthill}, \citenamefont {Ueda}, \citenamefont {Wang}, \citenamefont {Wang}, \citenamefont {Wang}, \citenamefont {Weber}, \citenamefont {Wolf}, \citenamefont {Young}, \citenamefont
  {Zhang}, \citenamefont {Bucksbaum}, \citenamefont {Marangos}, \citenamefont {Kling}, \citenamefont {Huang}, \citenamefont {Walter}, \citenamefont {Inhester}, \citenamefont {Berrah}, \citenamefont {Cryan},\ and\ \citenamefont {Marinelli}}]{Guo2024-ve}%
  \BibitemOpen
  \bibfield  {author} {\bibinfo {author} {\bibfnamefont {Z.}~\bibnamefont {Guo}}, \bibinfo {author} {\bibfnamefont {T.}~\bibnamefont {Driver}}, \bibinfo {author} {\bibfnamefont {S.}~\bibnamefont {Beauvarlet}}, \bibinfo {author} {\bibfnamefont {D.}~\bibnamefont {Cesar}}, \bibinfo {author} {\bibfnamefont {J.}~\bibnamefont {Duris}}, \bibinfo {author} {\bibfnamefont {P.~L.}\ \bibnamefont {Franz}}, \bibinfo {author} {\bibfnamefont {O.}~\bibnamefont {Alexander}}, \bibinfo {author} {\bibfnamefont {D.}~\bibnamefont {Bohler}}, \bibinfo {author} {\bibfnamefont {C.}~\bibnamefont {Bostedt}}, \bibinfo {author} {\bibfnamefont {V.}~\bibnamefont {Averbukh}}, \bibinfo {author} {\bibfnamefont {X.}~\bibnamefont {Cheng}}, \bibinfo {author} {\bibfnamefont {L.~F.}\ \bibnamefont {DiMauro}}, \bibinfo {author} {\bibfnamefont {G.}~\bibnamefont {Doumy}}, \bibinfo {author} {\bibfnamefont {R.}~\bibnamefont {Forbes}}, \bibinfo {author} {\bibfnamefont {O.}~\bibnamefont {Gessner}}, \bibinfo {author} {\bibfnamefont {J.~M.}\ \bibnamefont
  {Glownia}}, \bibinfo {author} {\bibfnamefont {E.}~\bibnamefont {Isele}}, \bibinfo {author} {\bibfnamefont {A.}~\bibnamefont {Kamalov}}, \bibinfo {author} {\bibfnamefont {K.~A.}\ \bibnamefont {Larsen}}, \bibinfo {author} {\bibfnamefont {S.}~\bibnamefont {Li}}, \bibinfo {author} {\bibfnamefont {X.}~\bibnamefont {Li}}, \bibinfo {author} {\bibfnamefont {M.-F.}\ \bibnamefont {Lin}}, \bibinfo {author} {\bibfnamefont {G.~A.}\ \bibnamefont {McCracken}}, \bibinfo {author} {\bibfnamefont {R.}~\bibnamefont {Obaid}}, \bibinfo {author} {\bibfnamefont {J.~T.}\ \bibnamefont {O'Neal}}, \bibinfo {author} {\bibfnamefont {R.~R.}\ \bibnamefont {Robles}}, \bibinfo {author} {\bibfnamefont {D.}~\bibnamefont {Rolles}}, \bibinfo {author} {\bibfnamefont {M.}~\bibnamefont {Ruberti}}, \bibinfo {author} {\bibfnamefont {A.}~\bibnamefont {Rudenko}}, \bibinfo {author} {\bibfnamefont {D.~S.}\ \bibnamefont {Slaughter}}, \bibinfo {author} {\bibfnamefont {N.~S.}\ \bibnamefont {Sudar}}, \bibinfo {author} {\bibfnamefont {E.}~\bibnamefont
  {Thierstein}}, \bibinfo {author} {\bibfnamefont {D.}~\bibnamefont {Tuthill}}, \bibinfo {author} {\bibfnamefont {K.}~\bibnamefont {Ueda}}, \bibinfo {author} {\bibfnamefont {E.}~\bibnamefont {Wang}}, \bibinfo {author} {\bibfnamefont {A.~L.}\ \bibnamefont {Wang}}, \bibinfo {author} {\bibfnamefont {J.}~\bibnamefont {Wang}}, \bibinfo {author} {\bibfnamefont {T.}~\bibnamefont {Weber}}, \bibinfo {author} {\bibfnamefont {T.~J.~A.}\ \bibnamefont {Wolf}}, \bibinfo {author} {\bibfnamefont {L.}~\bibnamefont {Young}}, \bibinfo {author} {\bibfnamefont {Z.}~\bibnamefont {Zhang}}, \bibinfo {author} {\bibfnamefont {P.~H.}\ \bibnamefont {Bucksbaum}}, \bibinfo {author} {\bibfnamefont {J.~P.}\ \bibnamefont {Marangos}}, \bibinfo {author} {\bibfnamefont {M.~F.}\ \bibnamefont {Kling}}, \bibinfo {author} {\bibfnamefont {Z.}~\bibnamefont {Huang}}, \bibinfo {author} {\bibfnamefont {P.}~\bibnamefont {Walter}}, \bibinfo {author} {\bibfnamefont {L.}~\bibnamefont {Inhester}}, \bibinfo {author} {\bibfnamefont {N.}~\bibnamefont {Berrah}},
  \bibinfo {author} {\bibfnamefont {J.~P.}\ \bibnamefont {Cryan}},\ and\ \bibinfo {author} {\bibfnamefont {A.}~\bibnamefont {Marinelli}},\ }\bibfield  {title} {\bibinfo {title} {Experimental demonstration of attosecond pump--probe spectroscopy with an x-ray free-electron laser},\ }\href@noop {} {\bibfield  {journal} {\bibinfo  {journal} {Nat. Photonics}\ }\textbf {\bibinfo {volume} {18}},\ \bibinfo {pages} {691} (\bibinfo {year} {2024})}\BibitemShut {NoStop}%
\bibitem [{\citenamefont {Budden}\ \emph {et~al.}(2021)\citenamefont {Budden}, \citenamefont {Gebert}, \citenamefont {Buzzi}, \citenamefont {Jotzu}, \citenamefont {Wang}, \citenamefont {Matsuyama}, \citenamefont {Meier}, \citenamefont {Laplace}, \citenamefont {Pontiroli}, \citenamefont {Ricc{\`o}} \emph {et~al.}}]{budden2021evidence}%
  \BibitemOpen
  \bibfield  {author} {\bibinfo {author} {\bibfnamefont {M.}~\bibnamefont {Budden}}, \bibinfo {author} {\bibfnamefont {T.}~\bibnamefont {Gebert}}, \bibinfo {author} {\bibfnamefont {M.}~\bibnamefont {Buzzi}}, \bibinfo {author} {\bibfnamefont {G.}~\bibnamefont {Jotzu}}, \bibinfo {author} {\bibfnamefont {E.}~\bibnamefont {Wang}}, \bibinfo {author} {\bibfnamefont {T.}~\bibnamefont {Matsuyama}}, \bibinfo {author} {\bibfnamefont {G.}~\bibnamefont {Meier}}, \bibinfo {author} {\bibfnamefont {Y.}~\bibnamefont {Laplace}}, \bibinfo {author} {\bibfnamefont {D.}~\bibnamefont {Pontiroli}}, \bibinfo {author} {\bibfnamefont {M.}~\bibnamefont {Ricc{\`o}}}, \emph {et~al.},\ }\bibfield  {title} {\bibinfo {title} {Evidence for metastable photo-induced superconductivity in k3c60},\ }\href@noop {} {\bibfield  {journal} {\bibinfo  {journal} {Nature Physics}\ }\textbf {\bibinfo {volume} {17}},\ \bibinfo {pages} {611} (\bibinfo {year} {2021})}\BibitemShut {NoStop}%
\bibitem [{\citenamefont {Rowe}\ \emph {et~al.}(2023)\citenamefont {Rowe}, \citenamefont {Yuan}, \citenamefont {Buzzi}, \citenamefont {Jotzu}, \citenamefont {Zhu}, \citenamefont {Fechner}, \citenamefont {F{\"o}rst}, \citenamefont {Liu}, \citenamefont {Pontiroli}, \citenamefont {Ricc{\`o}},\ and\ \citenamefont {Cavalleri}}]{Rowe2023-ab}%
  \BibitemOpen
  \bibfield  {author} {\bibinfo {author} {\bibfnamefont {E.}~\bibnamefont {Rowe}}, \bibinfo {author} {\bibfnamefont {B.}~\bibnamefont {Yuan}}, \bibinfo {author} {\bibfnamefont {M.}~\bibnamefont {Buzzi}}, \bibinfo {author} {\bibfnamefont {G.}~\bibnamefont {Jotzu}}, \bibinfo {author} {\bibfnamefont {Y.}~\bibnamefont {Zhu}}, \bibinfo {author} {\bibfnamefont {M.}~\bibnamefont {Fechner}}, \bibinfo {author} {\bibfnamefont {M.}~\bibnamefont {F{\"o}rst}}, \bibinfo {author} {\bibfnamefont {B.}~\bibnamefont {Liu}}, \bibinfo {author} {\bibfnamefont {D.}~\bibnamefont {Pontiroli}}, \bibinfo {author} {\bibfnamefont {M.}~\bibnamefont {Ricc{\`o}}},\ and\ \bibinfo {author} {\bibfnamefont {A.}~\bibnamefont {Cavalleri}},\ }\bibfield  {title} {\bibinfo {title} {Resonant enhancement of photo-induced superconductivity in {K3C60}},\ }\href@noop {} {\bibfield  {journal} {\bibinfo  {journal} {Nat. Phys.}\ }\textbf {\bibinfo {volume} {19}},\ \bibinfo {pages} {1821} (\bibinfo {year} {2023})}\BibitemShut {NoStop}%
\bibitem [{\citenamefont {Kemper}\ \emph {et~al.}(2015)\citenamefont {Kemper}, \citenamefont {Sentef}, \citenamefont {Moritz}, \citenamefont {Freericks},\ and\ \citenamefont {Devereaux}}]{PhysRevB.92.224517}%
  \BibitemOpen
  \bibfield  {author} {\bibinfo {author} {\bibfnamefont {A.~F.}\ \bibnamefont {Kemper}}, \bibinfo {author} {\bibfnamefont {M.~A.}\ \bibnamefont {Sentef}}, \bibinfo {author} {\bibfnamefont {B.}~\bibnamefont {Moritz}}, \bibinfo {author} {\bibfnamefont {J.~K.}\ \bibnamefont {Freericks}},\ and\ \bibinfo {author} {\bibfnamefont {T.~P.}\ \bibnamefont {Devereaux}},\ }\bibfield  {title} {\bibinfo {title} {Direct observation of higgs mode oscillations in the pump-probe photoemission spectra of electron-phonon mediated superconductors},\ }\href {https://doi.org/10.1103/PhysRevB.92.224517} {\bibfield  {journal} {\bibinfo  {journal} {Phys. Rev. B}\ }\textbf {\bibinfo {volume} {92}},\ \bibinfo {pages} {224517} (\bibinfo {year} {2015})}\BibitemShut {NoStop}%
\bibitem [{\citenamefont {Shimano}\ and\ \citenamefont {Tsuji}(2020)}]{higgs-mode-review}%
  \BibitemOpen
  \bibfield  {author} {\bibinfo {author} {\bibfnamefont {R.}~\bibnamefont {Shimano}}\ and\ \bibinfo {author} {\bibfnamefont {N.}~\bibnamefont {Tsuji}},\ }\bibfield  {title} {\bibinfo {title} {Higgs mode in superconductors},\ }\href {https://doi.org/https://doi.org/10.1146/annurev-conmatphys-031119-050813} {\bibfield  {journal} {\bibinfo  {journal} {Annual Review of Condensed Matter Physics}\ }\textbf {\bibinfo {volume} {11}},\ \bibinfo {pages} {103} (\bibinfo {year} {2020})}\BibitemShut {NoStop}%
\bibitem [{\citenamefont {Fausti}\ \emph {et~al.}(2011)\citenamefont {Fausti}, \citenamefont {Tobey}, \citenamefont {Dean}, \citenamefont {Kaiser}, \citenamefont {Dienst}, \citenamefont {Hoffmann}, \citenamefont {Pyon}, \citenamefont {Takayama}, \citenamefont {Takagi},\ and\ \citenamefont {Cavalleri}}]{fausti2011light}%
  \BibitemOpen
  \bibfield  {author} {\bibinfo {author} {\bibfnamefont {D.}~\bibnamefont {Fausti}}, \bibinfo {author} {\bibfnamefont {R.}~\bibnamefont {Tobey}}, \bibinfo {author} {\bibfnamefont {N.}~\bibnamefont {Dean}}, \bibinfo {author} {\bibfnamefont {S.}~\bibnamefont {Kaiser}}, \bibinfo {author} {\bibfnamefont {A.}~\bibnamefont {Dienst}}, \bibinfo {author} {\bibfnamefont {M.~C.}\ \bibnamefont {Hoffmann}}, \bibinfo {author} {\bibfnamefont {S.}~\bibnamefont {Pyon}}, \bibinfo {author} {\bibfnamefont {T.}~\bibnamefont {Takayama}}, \bibinfo {author} {\bibfnamefont {H.}~\bibnamefont {Takagi}},\ and\ \bibinfo {author} {\bibfnamefont {A.}~\bibnamefont {Cavalleri}},\ }\bibfield  {title} {\bibinfo {title} {Light-induced superconductivity in a stripe-ordered cuprate},\ }\href@noop {} {\bibfield  {journal} {\bibinfo  {journal} {science}\ }\textbf {\bibinfo {volume} {331}},\ \bibinfo {pages} {189} (\bibinfo {year} {2011})}\BibitemShut {NoStop}%
\bibitem [{\citenamefont {Von~Hoegen}\ \emph {et~al.}(2022)\citenamefont {Von~Hoegen}, \citenamefont {Fechner}, \citenamefont {F{\"o}rst}, \citenamefont {Taherian}, \citenamefont {Rowe}, \citenamefont {Ribak}, \citenamefont {Porras}, \citenamefont {Keimer}, \citenamefont {Michael}, \citenamefont {Demler} \emph {et~al.}}]{von2022amplification}%
  \BibitemOpen
  \bibfield  {author} {\bibinfo {author} {\bibfnamefont {A.}~\bibnamefont {Von~Hoegen}}, \bibinfo {author} {\bibfnamefont {M.}~\bibnamefont {Fechner}}, \bibinfo {author} {\bibfnamefont {M.}~\bibnamefont {F{\"o}rst}}, \bibinfo {author} {\bibfnamefont {N.}~\bibnamefont {Taherian}}, \bibinfo {author} {\bibfnamefont {E.}~\bibnamefont {Rowe}}, \bibinfo {author} {\bibfnamefont {A.}~\bibnamefont {Ribak}}, \bibinfo {author} {\bibfnamefont {J.}~\bibnamefont {Porras}}, \bibinfo {author} {\bibfnamefont {B.}~\bibnamefont {Keimer}}, \bibinfo {author} {\bibfnamefont {M.}~\bibnamefont {Michael}}, \bibinfo {author} {\bibfnamefont {E.}~\bibnamefont {Demler}}, \emph {et~al.},\ }\bibfield  {title} {\bibinfo {title} {Amplification of superconducting fluctuations in driven yba 2 cu 3 o 6+ x},\ }\href@noop {} {\bibfield  {journal} {\bibinfo  {journal} {Physical Review X}\ }\textbf {\bibinfo {volume} {12}},\ \bibinfo {pages} {031008} (\bibinfo {year} {2022})}\BibitemShut {NoStop}%
\bibitem [{\citenamefont {{Gol'tsman}}\ \emph {et~al.}(2001)\citenamefont {{Gol'tsman}}, \citenamefont {{Okunev}}, \citenamefont {{Chulkova}}, \citenamefont {{Lipatov}}, \citenamefont {{Semenov}}, \citenamefont {{Smirnov}}, \citenamefont {{Voronov}}, \citenamefont {{Dzardanov}}, \citenamefont {{Williams}},\ and\ \citenamefont {{Sobolewski}}}]{2001ApPhL..79..705G}%
  \BibitemOpen
  \bibfield  {author} {\bibinfo {author} {\bibfnamefont {G.~N.}\ \bibnamefont {{Gol'tsman}}}, \bibinfo {author} {\bibfnamefont {O.}~\bibnamefont {{Okunev}}}, \bibinfo {author} {\bibfnamefont {G.}~\bibnamefont {{Chulkova}}}, \bibinfo {author} {\bibfnamefont {A.}~\bibnamefont {{Lipatov}}}, \bibinfo {author} {\bibfnamefont {A.}~\bibnamefont {{Semenov}}}, \bibinfo {author} {\bibfnamefont {K.}~\bibnamefont {{Smirnov}}}, \bibinfo {author} {\bibfnamefont {B.}~\bibnamefont {{Voronov}}}, \bibinfo {author} {\bibfnamefont {A.}~\bibnamefont {{Dzardanov}}}, \bibinfo {author} {\bibfnamefont {C.}~\bibnamefont {{Williams}}},\ and\ \bibinfo {author} {\bibfnamefont {R.}~\bibnamefont {{Sobolewski}}},\ }\bibfield  {title} {\bibinfo {title} {{Picosecond superconducting single-photon optical detector}},\ }\href {https://doi.org/10.1063/1.1388868} {\bibfield  {journal} {\bibinfo  {journal} {Applied Physics Letters}\ }\textbf {\bibinfo {volume} {79}},\ \bibinfo {eid} {705} (\bibinfo {year} {2001})}\BibitemShut {NoStop}%
\bibitem [{\citenamefont {Simon}\ \emph {et~al.}(2025)\citenamefont {Simon}, \citenamefont {Foster}, \citenamefont {Sahoo}, \citenamefont {Shi}, \citenamefont {Batson}, \citenamefont {Incalza}, \citenamefont {Castellani}, \citenamefont {Medeiros}, \citenamefont {Heil},\ and\ \citenamefont {Berggren}}]{simon2025abinitiomodelingnonequilibrium}%
  \BibitemOpen
  \bibfield  {author} {\bibinfo {author} {\bibfnamefont {A.}~\bibnamefont {Simon}}, \bibinfo {author} {\bibfnamefont {R.}~\bibnamefont {Foster}}, \bibinfo {author} {\bibfnamefont {M.}~\bibnamefont {Sahoo}}, \bibinfo {author} {\bibfnamefont {J.}~\bibnamefont {Shi}}, \bibinfo {author} {\bibfnamefont {E.}~\bibnamefont {Batson}}, \bibinfo {author} {\bibfnamefont {F.}~\bibnamefont {Incalza}}, \bibinfo {author} {\bibfnamefont {M.}~\bibnamefont {Castellani}}, \bibinfo {author} {\bibfnamefont {O.}~\bibnamefont {Medeiros}}, \bibinfo {author} {\bibfnamefont {C.}~\bibnamefont {Heil}},\ and\ \bibinfo {author} {\bibfnamefont {K.~K.}\ \bibnamefont {Berggren}},\ }\bibfield  {title} {\bibinfo {title} {Ab initio modeling of nonequilibrium dynamics in superconducting detectors and qubits},\ }\href {https://doi.org/10.1103/3m2k-mzr6} {\bibfield  {journal} {\bibinfo  {journal} {Phys. Rev. B}\ }\textbf {\bibinfo {volume} {112}},\ \bibinfo {pages} {174512} (\bibinfo {year} {2025})}\BibitemShut {NoStop}%
\bibitem [{\citenamefont {Buzzi}\ \emph {et~al.}(2020)\citenamefont {Buzzi}, \citenamefont {Nicoletti}, \citenamefont {Fechner}, \citenamefont {Tancogne-Dejean}, \citenamefont {Sentef}, \citenamefont {Georges}, \citenamefont {Biesner}, \citenamefont {Uykur}, \citenamefont {Dressel}, \citenamefont {Henderson}, \citenamefont {Siegrist}, \citenamefont {Schlueter}, \citenamefont {Miyagawa}, \citenamefont {Kanoda}, \citenamefont {Nam}, \citenamefont {Ardavan}, \citenamefont {Coulthard}, \citenamefont {Tindall}, \citenamefont {Schlawin}, \citenamefont {Jaksch},\ and\ \citenamefont {Cavalleri}}]{PhysRevX.10.031028}%
  \BibitemOpen
  \bibfield  {author} {\bibinfo {author} {\bibfnamefont {M.}~\bibnamefont {Buzzi}}, \bibinfo {author} {\bibfnamefont {D.}~\bibnamefont {Nicoletti}}, \bibinfo {author} {\bibfnamefont {M.}~\bibnamefont {Fechner}}, \bibinfo {author} {\bibfnamefont {N.}~\bibnamefont {Tancogne-Dejean}}, \bibinfo {author} {\bibfnamefont {M.~A.}\ \bibnamefont {Sentef}}, \bibinfo {author} {\bibfnamefont {A.}~\bibnamefont {Georges}}, \bibinfo {author} {\bibfnamefont {T.}~\bibnamefont {Biesner}}, \bibinfo {author} {\bibfnamefont {E.}~\bibnamefont {Uykur}}, \bibinfo {author} {\bibfnamefont {M.}~\bibnamefont {Dressel}}, \bibinfo {author} {\bibfnamefont {A.}~\bibnamefont {Henderson}}, \bibinfo {author} {\bibfnamefont {T.}~\bibnamefont {Siegrist}}, \bibinfo {author} {\bibfnamefont {J.~A.}\ \bibnamefont {Schlueter}}, \bibinfo {author} {\bibfnamefont {K.}~\bibnamefont {Miyagawa}}, \bibinfo {author} {\bibfnamefont {K.}~\bibnamefont {Kanoda}}, \bibinfo {author} {\bibfnamefont {M.-S.}\ \bibnamefont {Nam}}, \bibinfo {author} {\bibfnamefont
  {A.}~\bibnamefont {Ardavan}}, \bibinfo {author} {\bibfnamefont {J.}~\bibnamefont {Coulthard}}, \bibinfo {author} {\bibfnamefont {J.}~\bibnamefont {Tindall}}, \bibinfo {author} {\bibfnamefont {F.}~\bibnamefont {Schlawin}}, \bibinfo {author} {\bibfnamefont {D.}~\bibnamefont {Jaksch}},\ and\ \bibinfo {author} {\bibfnamefont {A.}~\bibnamefont {Cavalleri}},\ }\bibfield  {title} {\bibinfo {title} {Photomolecular high-temperature superconductivity},\ }\href {https://doi.org/10.1103/PhysRevX.10.031028} {\bibfield  {journal} {\bibinfo  {journal} {Phys. Rev. X}\ }\textbf {\bibinfo {volume} {10}},\ \bibinfo {pages} {031028} (\bibinfo {year} {2020})}\BibitemShut {NoStop}%
\bibitem [{\citenamefont {Buzzi}\ \emph {et~al.}(2021)\citenamefont {Buzzi}, \citenamefont {Nicoletti}, \citenamefont {Fava}, \citenamefont {Jotzu}, \citenamefont {Miyagawa}, \citenamefont {Kanoda}, \citenamefont {Henderson}, \citenamefont {Siegrist}, \citenamefont {Schlueter}, \citenamefont {Nam}, \citenamefont {Ardavan},\ and\ \citenamefont {Cavalleri}}]{PhysRevLett.127.197002}%
  \BibitemOpen
  \bibfield  {author} {\bibinfo {author} {\bibfnamefont {M.}~\bibnamefont {Buzzi}}, \bibinfo {author} {\bibfnamefont {D.}~\bibnamefont {Nicoletti}}, \bibinfo {author} {\bibfnamefont {S.}~\bibnamefont {Fava}}, \bibinfo {author} {\bibfnamefont {G.}~\bibnamefont {Jotzu}}, \bibinfo {author} {\bibfnamefont {K.}~\bibnamefont {Miyagawa}}, \bibinfo {author} {\bibfnamefont {K.}~\bibnamefont {Kanoda}}, \bibinfo {author} {\bibfnamefont {A.}~\bibnamefont {Henderson}}, \bibinfo {author} {\bibfnamefont {T.}~\bibnamefont {Siegrist}}, \bibinfo {author} {\bibfnamefont {J.~A.}\ \bibnamefont {Schlueter}}, \bibinfo {author} {\bibfnamefont {M.-S.}\ \bibnamefont {Nam}}, \bibinfo {author} {\bibfnamefont {A.}~\bibnamefont {Ardavan}},\ and\ \bibinfo {author} {\bibfnamefont {A.}~\bibnamefont {Cavalleri}},\ }\bibfield  {title} {\bibinfo {title} {Phase diagram for light-induced superconductivity in $\ensuremath{\kappa}\text{\ensuremath{-}}(\mathrm{ET}{)}_{2}\text{\ensuremath{-}}\mathrm{X}$},\ }\href
  {https://doi.org/10.1103/PhysRevLett.127.197002} {\bibfield  {journal} {\bibinfo  {journal} {Phys. Rev. Lett.}\ }\textbf {\bibinfo {volume} {127}},\ \bibinfo {pages} {197002} (\bibinfo {year} {2021})}\BibitemShut {NoStop}%
\bibitem [{\citenamefont {Prasankumar}\ \emph {et~al.}(2026)\citenamefont {Prasankumar}, \citenamefont {Julian}, \citenamefont {Hutcheon}, \citenamefont {Heil}, \citenamefont {Deng}, \citenamefont {Basov}, \citenamefont {Chu}, \citenamefont {Comin}, \citenamefont {Kim}, \citenamefont {Meredig}, \citenamefont {Pickard}, \citenamefont {Pickett}, \citenamefont {Strobel}, \citenamefont {Wolf}, \citenamefont {Zurek},\ and\ \citenamefont {Myhrvold}}]{doi:10.1073/pnas.2520324123}%
  \BibitemOpen
  \bibfield  {author} {\bibinfo {author} {\bibfnamefont {R.~P.}\ \bibnamefont {Prasankumar}}, \bibinfo {author} {\bibfnamefont {M.}~\bibnamefont {Julian}}, \bibinfo {author} {\bibfnamefont {M.}~\bibnamefont {Hutcheon}}, \bibinfo {author} {\bibfnamefont {C.}~\bibnamefont {Heil}}, \bibinfo {author} {\bibfnamefont {L.}~\bibnamefont {Deng}}, \bibinfo {author} {\bibfnamefont {D.}~\bibnamefont {Basov}}, \bibinfo {author} {\bibfnamefont {C.-W.}\ \bibnamefont {Chu}}, \bibinfo {author} {\bibfnamefont {R.}~\bibnamefont {Comin}}, \bibinfo {author} {\bibfnamefont {P.}~\bibnamefont {Kim}}, \bibinfo {author} {\bibfnamefont {B.}~\bibnamefont {Meredig}}, \bibinfo {author} {\bibfnamefont {C.}~\bibnamefont {Pickard}}, \bibinfo {author} {\bibfnamefont {W.~E.}\ \bibnamefont {Pickett}}, \bibinfo {author} {\bibfnamefont {T.}~\bibnamefont {Strobel}}, \bibinfo {author} {\bibfnamefont {S.}~\bibnamefont {Wolf}}, \bibinfo {author} {\bibfnamefont {E.}~\bibnamefont {Zurek}},\ and\ \bibinfo {author} {\bibfnamefont {N.}~\bibnamefont
  {Myhrvold}},\ }\bibfield  {title} {\bibinfo {title} {The path to room-temperature superconductivity: A programmatic approach},\ }\href {https://doi.org/10.1073/pnas.2520324123} {\bibfield  {journal} {\bibinfo  {journal} {Proceedings of the National Academy of Sciences}\ }\textbf {\bibinfo {volume} {123}},\ \bibinfo {pages} {e2520324123} (\bibinfo {year} {2026})},\ \Eprint {https://arxiv.org/abs/https://www.pnas.org/doi/pdf/10.1073/pnas.2520324123} {https://www.pnas.org/doi/pdf/10.1073/pnas.2520324123} \BibitemShut {NoStop}%
\bibitem [{\citenamefont {Komnik}\ and\ \citenamefont {Thorwart}(2016)}]{Komnik-2016}%
  \BibitemOpen
  \bibfield  {author} {\bibinfo {author} {\bibfnamefont {A.}~\bibnamefont {Komnik}}\ and\ \bibinfo {author} {\bibfnamefont {M.}~\bibnamefont {Thorwart}},\ }\bibfield  {title} {\bibinfo {title} {Bcs theory of driven superconductivity},\ }\href@noop {} {\bibfield  {journal} {\bibinfo  {journal} {Eur. Phys. J. B}\ }\textbf {\bibinfo {volume} {89}} (\bibinfo {year} {2016})}\BibitemShut {NoStop}%
\bibitem [{\citenamefont {Knap}\ \emph {et~al.}(2016)\citenamefont {Knap}, \citenamefont {Babadi}, \citenamefont {Refael}, \citenamefont {Martin},\ and\ \citenamefont {Demler}}]{PhysRevB.94.214504}%
  \BibitemOpen
  \bibfield  {author} {\bibinfo {author} {\bibfnamefont {M.}~\bibnamefont {Knap}}, \bibinfo {author} {\bibfnamefont {M.}~\bibnamefont {Babadi}}, \bibinfo {author} {\bibfnamefont {G.}~\bibnamefont {Refael}}, \bibinfo {author} {\bibfnamefont {I.}~\bibnamefont {Martin}},\ and\ \bibinfo {author} {\bibfnamefont {E.}~\bibnamefont {Demler}},\ }\bibfield  {title} {\bibinfo {title} {Dynamical cooper pairing in nonequilibrium electron-phonon systems},\ }\href {https://doi.org/10.1103/PhysRevB.94.214504} {\bibfield  {journal} {\bibinfo  {journal} {Phys. Rev. B}\ }\textbf {\bibinfo {volume} {94}},\ \bibinfo {pages} {214504} (\bibinfo {year} {2016})}\BibitemShut {NoStop}%
\bibitem [{\citenamefont {Kim}\ \emph {et~al.}(2016)\citenamefont {Kim}, \citenamefont {Nomura}, \citenamefont {Ferrero}, \citenamefont {Seth}, \citenamefont {Parcollet},\ and\ \citenamefont {Georges}}]{PhysRevB.94.155152}%
  \BibitemOpen
  \bibfield  {author} {\bibinfo {author} {\bibfnamefont {M.}~\bibnamefont {Kim}}, \bibinfo {author} {\bibfnamefont {Y.}~\bibnamefont {Nomura}}, \bibinfo {author} {\bibfnamefont {M.}~\bibnamefont {Ferrero}}, \bibinfo {author} {\bibfnamefont {P.}~\bibnamefont {Seth}}, \bibinfo {author} {\bibfnamefont {O.}~\bibnamefont {Parcollet}},\ and\ \bibinfo {author} {\bibfnamefont {A.}~\bibnamefont {Georges}},\ }\bibfield  {title} {\bibinfo {title} {Enhancing superconductivity in ${A}_{3}{\mathrm{c}}_{60}$ fullerides},\ }\href {https://doi.org/10.1103/PhysRevB.94.155152} {\bibfield  {journal} {\bibinfo  {journal} {Phys. Rev. B}\ }\textbf {\bibinfo {volume} {94}},\ \bibinfo {pages} {155152} (\bibinfo {year} {2016})}\BibitemShut {NoStop}%
\bibitem [{\citenamefont {Sentef}\ \emph {et~al.}(2016)\citenamefont {Sentef}, \citenamefont {Kemper}, \citenamefont {Georges},\ and\ \citenamefont {Kollath}}]{PhysRevB.93.144506}%
  \BibitemOpen
  \bibfield  {author} {\bibinfo {author} {\bibfnamefont {M.~A.}\ \bibnamefont {Sentef}}, \bibinfo {author} {\bibfnamefont {A.~F.}\ \bibnamefont {Kemper}}, \bibinfo {author} {\bibfnamefont {A.}~\bibnamefont {Georges}},\ and\ \bibinfo {author} {\bibfnamefont {C.}~\bibnamefont {Kollath}},\ }\bibfield  {title} {\bibinfo {title} {Theory of light-enhanced phonon-mediated superconductivity},\ }\href {https://doi.org/10.1103/PhysRevB.93.144506} {\bibfield  {journal} {\bibinfo  {journal} {Phys. Rev. B}\ }\textbf {\bibinfo {volume} {93}},\ \bibinfo {pages} {144506} (\bibinfo {year} {2016})}\BibitemShut {NoStop}%
\bibitem [{\citenamefont {Ido}\ \emph {et~al.}(2017)\citenamefont {Ido}, \citenamefont {Ohgoe},\ and\ \citenamefont {Imada}}]{doi:10.1126/sciadv.1700718}%
  \BibitemOpen
  \bibfield  {author} {\bibinfo {author} {\bibfnamefont {K.}~\bibnamefont {Ido}}, \bibinfo {author} {\bibfnamefont {T.}~\bibnamefont {Ohgoe}},\ and\ \bibinfo {author} {\bibfnamefont {M.}~\bibnamefont {Imada}},\ }\bibfield  {title} {\bibinfo {title} {Correlation-induced superconductivity dynamically stabilized and enhanced by laser irradiation},\ }\href {https://doi.org/10.1126/sciadv.1700718} {\bibfield  {journal} {\bibinfo  {journal} {Science Advances}\ }\textbf {\bibinfo {volume} {3}},\ \bibinfo {pages} {e1700718} (\bibinfo {year} {2017})},\ \Eprint {https://arxiv.org/abs/https://www.science.org/doi/pdf/10.1126/sciadv.1700718} {https://www.science.org/doi/pdf/10.1126/sciadv.1700718} \BibitemShut {NoStop}%
\bibitem [{\citenamefont {Kennes}\ \emph {et~al.}(2017)\citenamefont {Kennes}, \citenamefont {Wilner}, \citenamefont {Reichman},\ and\ \citenamefont {Millis}}]{Kennes2017-qo}%
  \BibitemOpen
  \bibfield  {author} {\bibinfo {author} {\bibfnamefont {D.~M.}\ \bibnamefont {Kennes}}, \bibinfo {author} {\bibfnamefont {E.~Y.}\ \bibnamefont {Wilner}}, \bibinfo {author} {\bibfnamefont {D.~R.}\ \bibnamefont {Reichman}},\ and\ \bibinfo {author} {\bibfnamefont {A.~J.}\ \bibnamefont {Millis}},\ }\bibfield  {title} {\bibinfo {title} {Transient superconductivity from electronic squeezing of optically pumped phonons},\ }\href@noop {} {\bibfield  {journal} {\bibinfo  {journal} {Nat. Phys.}\ }\textbf {\bibinfo {volume} {13}},\ \bibinfo {pages} {479} (\bibinfo {year} {2017})}\BibitemShut {NoStop}%
\bibitem [{\citenamefont {Dasari}\ and\ \citenamefont {Eckstein}(2018)}]{PhysRevB.98.235149}%
  \BibitemOpen
  \bibfield  {author} {\bibinfo {author} {\bibfnamefont {N.}~\bibnamefont {Dasari}}\ and\ \bibinfo {author} {\bibfnamefont {M.}~\bibnamefont {Eckstein}},\ }\bibfield  {title} {\bibinfo {title} {Transient floquet engineering of superconductivity},\ }\href {https://doi.org/10.1103/PhysRevB.98.235149} {\bibfield  {journal} {\bibinfo  {journal} {Phys. Rev. B}\ }\textbf {\bibinfo {volume} {98}},\ \bibinfo {pages} {235149} (\bibinfo {year} {2018})}\BibitemShut {NoStop}%
\bibitem [{\citenamefont {Nava}\ \emph {et~al.}(2018)\citenamefont {Nava}, \citenamefont {Giannetti}, \citenamefont {Georges}, \citenamefont {Tosatti},\ and\ \citenamefont {Fabrizio}}]{Nava2018-ky}%
  \BibitemOpen
  \bibfield  {author} {\bibinfo {author} {\bibfnamefont {A.}~\bibnamefont {Nava}}, \bibinfo {author} {\bibfnamefont {C.}~\bibnamefont {Giannetti}}, \bibinfo {author} {\bibfnamefont {A.}~\bibnamefont {Georges}}, \bibinfo {author} {\bibfnamefont {E.}~\bibnamefont {Tosatti}},\ and\ \bibinfo {author} {\bibfnamefont {M.}~\bibnamefont {Fabrizio}},\ }\bibfield  {title} {\bibinfo {title} {Cooling quasiparticles in {A3C60} fullerides by excitonic mid-infrared absorption},\ }\href@noop {} {\bibfield  {journal} {\bibinfo  {journal} {Nat. Phys.}\ }\textbf {\bibinfo {volume} {14}},\ \bibinfo {pages} {154} (\bibinfo {year} {2018})}\BibitemShut {NoStop}%
\bibitem [{\citenamefont {Chattopadhyay}\ \emph {et~al.}(2026)\citenamefont {Chattopadhyay}, \citenamefont {Michael}, \citenamefont {Cavalleri},\ and\ \citenamefont {Demler}}]{chattopadhyay2026giant}%
  \BibitemOpen
  \bibfield  {author} {\bibinfo {author} {\bibfnamefont {S.}~\bibnamefont {Chattopadhyay}}, \bibinfo {author} {\bibfnamefont {M.}~\bibnamefont {Michael}}, \bibinfo {author} {\bibfnamefont {A.}~\bibnamefont {Cavalleri}},\ and\ \bibinfo {author} {\bibfnamefont {E.}~\bibnamefont {Demler}},\ }\bibfield  {title} {\bibinfo {title} {Giant resonant enhancement of photoinduced dynamical cooper pairing, far above $ t\_c$},\ }\href@noop {} {\bibfield  {journal} {\bibinfo  {journal} {arXiv preprint arXiv:2601.18712}\ } (\bibinfo {year} {2026})}\BibitemShut {NoStop}%
\bibitem [{\citenamefont {Chang}\ and\ \citenamefont {Scalapino}(1977)}]{chang1977kinetic}%
  \BibitemOpen
  \bibfield  {author} {\bibinfo {author} {\bibfnamefont {J.-J.}\ \bibnamefont {Chang}}\ and\ \bibinfo {author} {\bibfnamefont {D.}~\bibnamefont {Scalapino}},\ }\bibfield  {title} {\bibinfo {title} {Kinetic-equation approach to nonequilibrium superconductivity},\ }\href@noop {} {\bibfield  {journal} {\bibinfo  {journal} {Physical Review B}\ }\textbf {\bibinfo {volume} {15}},\ \bibinfo {pages} {2651} (\bibinfo {year} {1977})}\BibitemShut {NoStop}%
\bibitem [{\citenamefont {Simon}(2025)}]{simon2025ab}%
  \BibitemOpen
  \bibfield  {author} {\bibinfo {author} {\bibfnamefont {A.}~\bibnamefont {Simon}},\ }\emph {\bibinfo {title} {Ab initio modeling of superconducting nanowire single-photon detectors}},\ \href@noop {} {Master's thesis},\ \bibinfo  {school} {Massachusetts Institute of Technology} (\bibinfo {year} {2025})\BibitemShut {NoStop}%
\bibitem [{\citenamefont {Kaplan}\ \emph {et~al.}(1976)\citenamefont {Kaplan}, \citenamefont {Chi}, \citenamefont {Langenberg}, \citenamefont {Chang}, \citenamefont {Jafarey},\ and\ \citenamefont {Scalapino}}]{kaplan1976quasiparticle}%
  \BibitemOpen
  \bibfield  {author} {\bibinfo {author} {\bibfnamefont {S.~B.}\ \bibnamefont {Kaplan}}, \bibinfo {author} {\bibfnamefont {C.}~\bibnamefont {Chi}}, \bibinfo {author} {\bibfnamefont {D.}~\bibnamefont {Langenberg}}, \bibinfo {author} {\bibfnamefont {J.-J.}\ \bibnamefont {Chang}}, \bibinfo {author} {\bibfnamefont {S.}~\bibnamefont {Jafarey}},\ and\ \bibinfo {author} {\bibfnamefont {D.}~\bibnamefont {Scalapino}},\ }\bibfield  {title} {\bibinfo {title} {Quasiparticle and phonon lifetimes in superconductors},\ }\href@noop {} {\bibfield  {journal} {\bibinfo  {journal} {Physical Review B}\ }\textbf {\bibinfo {volume} {14}},\ \bibinfo {pages} {4854} (\bibinfo {year} {1976})}\BibitemShut {NoStop}%
\bibitem [{\citenamefont {Kadanoff}(2018)}]{kadanoff2018quantum}%
  \BibitemOpen
  \bibfield  {author} {\bibinfo {author} {\bibfnamefont {B.~G.}\ \bibnamefont {Kadanoff}, \bibfnamefont {Leo~P}},\ }\href@noop {} {\emph {\bibinfo {title} {Quantum statistical mechanics}}}\ (\bibinfo  {publisher} {CRC Press},\ \bibinfo {year} {2018})\BibitemShut {NoStop}%
\bibitem [{\citenamefont {Prange}\ and\ \citenamefont {Kadanoff}(1964)}]{prange1964transport}%
  \BibitemOpen
  \bibfield  {author} {\bibinfo {author} {\bibfnamefont {R.~E.}\ \bibnamefont {Prange}}\ and\ \bibinfo {author} {\bibfnamefont {L.~P.}\ \bibnamefont {Kadanoff}},\ }\bibfield  {title} {\bibinfo {title} {Transport theory for electron-phonon interactions in metals},\ }\href@noop {} {\bibfield  {journal} {\bibinfo  {journal} {Physical Review}\ }\textbf {\bibinfo {volume} {134}},\ \bibinfo {pages} {A566} (\bibinfo {year} {1964})}\BibitemShut {NoStop}%
\bibitem [{See supplemental information()}]{seeSI}%
  \BibitemOpen
  See supplemental information,\ \href@noop {} {}\BibitemShut {NoStop}%
\bibitem [{\citenamefont {Little}(1959)}]{little1959transport}%
  \BibitemOpen
  \bibfield  {author} {\bibinfo {author} {\bibfnamefont {W.}~\bibnamefont {Little}},\ }\bibfield  {title} {\bibinfo {title} {The transport of heat between dissimilar solids at low temperatures},\ }\href@noop {} {\bibfield  {journal} {\bibinfo  {journal} {Canadian Journal of Physics}\ }\textbf {\bibinfo {volume} {37}},\ \bibinfo {pages} {334} (\bibinfo {year} {1959})}\BibitemShut {NoStop}%
\bibitem [{\citenamefont {Simon}\ \emph {et~al.}(2026)\citenamefont {Simon}, \citenamefont {Shi}, \citenamefont {Spath}, \citenamefont {Kogler}, \citenamefont {Foster}, \citenamefont {Batson}, \citenamefont {Ferreira}, \citenamefont {Sahoo}, \citenamefont {Keathley}, \citenamefont {Pickett}, \citenamefont {Prasankumar}, \citenamefont {Berggren},\ and\ \citenamefont {Heil}}]{FIXME-THIS-PAPER}%
  \BibitemOpen
  \bibfield  {author} {\bibinfo {author} {\bibfnamefont {A.}~\bibnamefont {Simon}}, \bibinfo {author} {\bibfnamefont {J.}~\bibnamefont {Shi}}, \bibinfo {author} {\bibfnamefont {D.}~\bibnamefont {Spath}}, \bibinfo {author} {\bibfnamefont {E.}~\bibnamefont {Kogler}}, \bibinfo {author} {\bibfnamefont {R.}~\bibnamefont {Foster}}, \bibinfo {author} {\bibfnamefont {E.}~\bibnamefont {Batson}}, \bibinfo {author} {\bibfnamefont {P.~N.}\ \bibnamefont {Ferreira}}, \bibinfo {author} {\bibfnamefont {M.}~\bibnamefont {Sahoo}}, \bibinfo {author} {\bibfnamefont {P.~D.}\ \bibnamefont {Keathley}}, \bibinfo {author} {\bibfnamefont {W.~E.}\ \bibnamefont {Pickett}}, \bibinfo {author} {\bibfnamefont {R.}~\bibnamefont {Prasankumar}}, \bibinfo {author} {\bibfnamefont {K.~K.}\ \bibnamefont {Berggren}},\ and\ \bibinfo {author} {\bibfnamefont {C.}~\bibnamefont {Heil}},\ }\bibfield  {title} {\bibinfo {title} {Fast real-axis eliashberg calculations: Full-bandwidth solutions beyond the constant density of states approximation},\
  }\href@noop {} {\bibfield  {journal} {\bibinfo  {journal} {arXiv}\ } (\bibinfo {year} {2026})}\BibitemShut {NoStop}%
\bibitem [{\citenamefont {Kogler}\ \emph {et~al.}(2025)\citenamefont {Kogler}, \citenamefont {Spath}, \citenamefont {Lucrezi}, \citenamefont {Mori}, \citenamefont {Zhu}, \citenamefont {Li}, \citenamefont {Margine},\ and\ \citenamefont {Heil}}]{IsoME}%
  \BibitemOpen
  \bibfield  {author} {\bibinfo {author} {\bibfnamefont {E.}~\bibnamefont {Kogler}}, \bibinfo {author} {\bibfnamefont {D.}~\bibnamefont {Spath}}, \bibinfo {author} {\bibfnamefont {R.}~\bibnamefont {Lucrezi}}, \bibinfo {author} {\bibfnamefont {H.}~\bibnamefont {Mori}}, \bibinfo {author} {\bibfnamefont {Z.}~\bibnamefont {Zhu}}, \bibinfo {author} {\bibfnamefont {Z.}~\bibnamefont {Li}}, \bibinfo {author} {\bibfnamefont {E.~R.}\ \bibnamefont {Margine}},\ and\ \bibinfo {author} {\bibfnamefont {C.}~\bibnamefont {Heil}},\ }\href {https://arxiv.org/abs/2503.03559} {\bibinfo {title} {Isome: Streamlining high-precision eliashberg calculations}} (\bibinfo {year} {2025}),\ \Eprint {https://arxiv.org/abs/2503.03559} {arXiv:2503.03559 [cond-mat.supr-con]} \BibitemShut {NoStop}%
\bibitem [{\citenamefont {Ponc{\'e}}\ \emph {et~al.}(2016)\citenamefont {Ponc{\'e}}, \citenamefont {Margine}, \citenamefont {Verdi},\ and\ \citenamefont {Giustino}}]{ponce2016epw}%
  \BibitemOpen
  \bibfield  {author} {\bibinfo {author} {\bibfnamefont {S.}~\bibnamefont {Ponc{\'e}}}, \bibinfo {author} {\bibfnamefont {E.~R.}\ \bibnamefont {Margine}}, \bibinfo {author} {\bibfnamefont {C.}~\bibnamefont {Verdi}},\ and\ \bibinfo {author} {\bibfnamefont {F.}~\bibnamefont {Giustino}},\ }\bibfield  {title} {\bibinfo {title} {Epw: Electron--phonon coupling, transport and superconducting properties using maximally localized wannier functions},\ }\href@noop {} {\bibfield  {journal} {\bibinfo  {journal} {Computer Physics Communications}\ }\textbf {\bibinfo {volume} {209}},\ \bibinfo {pages} {116} (\bibinfo {year} {2016})}\BibitemShut {NoStop}%
\bibitem [{\citenamefont {Kraberger}\ \emph {et~al.}(2017)\citenamefont {Kraberger}, \citenamefont {Triebl}, \citenamefont {Zingl},\ and\ \citenamefont {Aichhorn}}]{kraberger2017maximum}%
  \BibitemOpen
  \bibfield  {author} {\bibinfo {author} {\bibfnamefont {G.~J.}\ \bibnamefont {Kraberger}}, \bibinfo {author} {\bibfnamefont {R.}~\bibnamefont {Triebl}}, \bibinfo {author} {\bibfnamefont {M.}~\bibnamefont {Zingl}},\ and\ \bibinfo {author} {\bibfnamefont {M.}~\bibnamefont {Aichhorn}},\ }\bibfield  {title} {\bibinfo {title} {Maximum entropy formalism for the analytic continuation of matrix-valued green's functions},\ }\href@noop {} {\bibfield  {journal} {\bibinfo  {journal} {Physical Review B}\ }\textbf {\bibinfo {volume} {96}},\ \bibinfo {pages} {155128} (\bibinfo {year} {2017})}\BibitemShut {NoStop}%
\bibitem [{\citenamefont {Beach}\ \emph {et~al.}(2000)\citenamefont {Beach}, \citenamefont {Gooding},\ and\ \citenamefont {Marsiglio}}]{PhysRevB.61.5147}%
  \BibitemOpen
  \bibfield  {author} {\bibinfo {author} {\bibfnamefont {K.~S.~D.}\ \bibnamefont {Beach}}, \bibinfo {author} {\bibfnamefont {R.~J.}\ \bibnamefont {Gooding}},\ and\ \bibinfo {author} {\bibfnamefont {F.}~\bibnamefont {Marsiglio}},\ }\bibfield  {title} {\bibinfo {title} {Reliable pad\'e analytical continuation method based on a high-accuracy symbolic computation algorithm},\ }\href {https://doi.org/10.1103/PhysRevB.61.5147} {\bibfield  {journal} {\bibinfo  {journal} {Phys. Rev. B}\ }\textbf {\bibinfo {volume} {61}},\ \bibinfo {pages} {5147} (\bibinfo {year} {2000})}\BibitemShut {NoStop}%
\bibitem [{\citenamefont {Khodachenko}\ \emph {et~al.}(2024)\citenamefont {Khodachenko}, \citenamefont {Lucrezi}, \citenamefont {Ferreira}, \citenamefont {Aichhorn},\ and\ \citenamefont {Heil}}]{khodachenko2024nevanlinna}%
  \BibitemOpen
  \bibfield  {author} {\bibinfo {author} {\bibfnamefont {D.}~\bibnamefont {Khodachenko}}, \bibinfo {author} {\bibfnamefont {R.}~\bibnamefont {Lucrezi}}, \bibinfo {author} {\bibfnamefont {P.}~\bibnamefont {Ferreira}}, \bibinfo {author} {\bibfnamefont {M.}~\bibnamefont {Aichhorn}},\ and\ \bibinfo {author} {\bibfnamefont {C.}~\bibnamefont {Heil}},\ }\bibfield  {title} {\bibinfo {title} {Nevanlinna analytic continuation for migdal--eliashberg theory},\ }\href {https://doi.org/https://doi.org/10.1016/j.commt.2024.100015} {\bibfield  {journal} {\bibinfo  {journal} {Computational Materials Today}\ }\textbf {\bibinfo {volume} {4}},\ \bibinfo {pages} {100015} (\bibinfo {year} {2024})}\BibitemShut {NoStop}%
\bibitem [{\citenamefont {Lobo}\ \emph {et~al.}(2005)\citenamefont {Lobo}, \citenamefont {LaVeigne}, \citenamefont {Reitze}, \citenamefont {Tanner}, \citenamefont {Barber}, \citenamefont {Jacques}, \citenamefont {Bosland}, \citenamefont {Burns},\ and\ \citenamefont {Carr}}]{PhysRevB.72.024510}%
  \BibitemOpen
  \bibfield  {author} {\bibinfo {author} {\bibfnamefont {R.~P. S.~M.}\ \bibnamefont {Lobo}}, \bibinfo {author} {\bibfnamefont {J.~D.}\ \bibnamefont {LaVeigne}}, \bibinfo {author} {\bibfnamefont {D.~H.}\ \bibnamefont {Reitze}}, \bibinfo {author} {\bibfnamefont {D.~B.}\ \bibnamefont {Tanner}}, \bibinfo {author} {\bibfnamefont {Z.~H.}\ \bibnamefont {Barber}}, \bibinfo {author} {\bibfnamefont {E.}~\bibnamefont {Jacques}}, \bibinfo {author} {\bibfnamefont {P.}~\bibnamefont {Bosland}}, \bibinfo {author} {\bibfnamefont {M.~J.}\ \bibnamefont {Burns}},\ and\ \bibinfo {author} {\bibfnamefont {G.~L.}\ \bibnamefont {Carr}},\ }\bibfield  {title} {\bibinfo {title} {Photoinduced time-resolved electrodynamics of superconducting metals and alloys},\ }\href {https://doi.org/10.1103/PhysRevB.72.024510} {\bibfield  {journal} {\bibinfo  {journal} {Phys. Rev. B}\ }\textbf {\bibinfo {volume} {72}},\ \bibinfo {pages} {024510} (\bibinfo {year} {2005})}\BibitemShut {NoStop}%
\bibitem [{\citenamefont {Wu}\ \emph {et~al.}(2024)\citenamefont {Wu}, \citenamefont {Yu}, \citenamefont {Hasaien}, \citenamefont {Hong}, \citenamefont {Shan}, \citenamefont {Tian}, \citenamefont {Zhai}, \citenamefont {Hu}, \citenamefont {Cheng},\ and\ \citenamefont {Zhao}}]{Wu2024-rm}%
  \BibitemOpen
  \bibfield  {author} {\bibinfo {author} {\bibfnamefont {Y.~L.}\ \bibnamefont {Wu}}, \bibinfo {author} {\bibfnamefont {X.~H.}\ \bibnamefont {Yu}}, \bibinfo {author} {\bibfnamefont {J.~Z.~L.}\ \bibnamefont {Hasaien}}, \bibinfo {author} {\bibfnamefont {F.}~\bibnamefont {Hong}}, \bibinfo {author} {\bibfnamefont {P.~F.}\ \bibnamefont {Shan}}, \bibinfo {author} {\bibfnamefont {Z.~Y.}\ \bibnamefont {Tian}}, \bibinfo {author} {\bibfnamefont {Y.~N.}\ \bibnamefont {Zhai}}, \bibinfo {author} {\bibfnamefont {J.~P.}\ \bibnamefont {Hu}}, \bibinfo {author} {\bibfnamefont {J.~G.}\ \bibnamefont {Cheng}},\ and\ \bibinfo {author} {\bibfnamefont {J.}~\bibnamefont {Zhao}},\ }\bibfield  {title} {\bibinfo {title} {Ultrafast dynamics evidence of strong coupling superconductivity in {LaH10$\pm$$\delta$}},\ }\href@noop {} {\bibfield  {journal} {\bibinfo  {journal} {Nat. Commun.}\ }\textbf {\bibinfo {volume} {15}},\ \bibinfo {pages} {9683} (\bibinfo {year} {2024})}\BibitemShut {NoStop}%
\bibitem [{\citenamefont {Mahan}(1980)}]{mahan2013many}%
  \BibitemOpen
  \bibfield  {author} {\bibinfo {author} {\bibfnamefont {G.~D.}\ \bibnamefont {Mahan}},\ }\href@noop {} {\emph {\bibinfo {title} {Many-particle physics}}}\ (\bibinfo  {publisher} {Springer Science \& Business Media},\ \bibinfo {year} {1980})\BibitemShut {NoStop}%
\bibitem [{\citenamefont {Nam}(1967)}]{nam1967theory}%
  \BibitemOpen
  \bibfield  {author} {\bibinfo {author} {\bibfnamefont {S.~B.}\ \bibnamefont {Nam}},\ }\bibfield  {title} {\bibinfo {title} {Theory of electromagnetic properties of strong-coupling and impure superconductors. ii},\ }\href@noop {} {\bibfield  {journal} {\bibinfo  {journal} {Physical Review}\ }\textbf {\bibinfo {volume} {156}},\ \bibinfo {pages} {487} (\bibinfo {year} {1967})}\BibitemShut {NoStop}%
\bibitem [{\citenamefont {Dodge}\ \emph {et~al.}(2023)\citenamefont {Dodge}, \citenamefont {Lopez},\ and\ \citenamefont {Sahota}}]{PhysRevLett.130.146002}%
  \BibitemOpen
  \bibfield  {author} {\bibinfo {author} {\bibfnamefont {J.~S.}\ \bibnamefont {Dodge}}, \bibinfo {author} {\bibfnamefont {L.}~\bibnamefont {Lopez}},\ and\ \bibinfo {author} {\bibfnamefont {D.~G.}\ \bibnamefont {Sahota}},\ }\bibfield  {title} {\bibinfo {title} {Optical saturation produces spurious evidence for photoinduced superconductivity in ${\mathrm{k}}_{3}{\mathrm{c}}_{60}$},\ }\href {https://doi.org/10.1103/PhysRevLett.130.146002} {\bibfield  {journal} {\bibinfo  {journal} {Phys. Rev. Lett.}\ }\textbf {\bibinfo {volume} {130}},\ \bibinfo {pages} {146002} (\bibinfo {year} {2023})}\BibitemShut {NoStop}%
\bibitem [{\citenamefont {Giustino}(2017)}]{giustino2017electron}%
  \BibitemOpen
  \bibfield  {author} {\bibinfo {author} {\bibfnamefont {F.}~\bibnamefont {Giustino}},\ }\bibfield  {title} {\bibinfo {title} {Electron-phonon interactions from first principles},\ }\href@noop {} {\bibfield  {journal} {\bibinfo  {journal} {Reviews of Modern Physics}\ }\textbf {\bibinfo {volume} {89}},\ \bibinfo {pages} {015003} (\bibinfo {year} {2017})}\BibitemShut {NoStop}%
\bibitem [{\citenamefont {Giannozzi}\ \emph {et~al.}(2017)\citenamefont {Giannozzi}, \citenamefont {Andreussi}, \citenamefont {Brumme}, \citenamefont {Bunau}, \citenamefont {Nardelli}, \citenamefont {Calandra}, \citenamefont {Car}, \citenamefont {Cavazzoni}, \citenamefont {Ceresoli}, \citenamefont {Cococcioni}, \citenamefont {Colonna}, \citenamefont {Carnimeo}, \citenamefont {Corso}, \citenamefont {de~Gironcoli}, \citenamefont {Delugas}, \citenamefont {Jr}, \citenamefont {Ferretti}, \citenamefont {Floris}, \citenamefont {Fratesi}, \citenamefont {Fugallo}, \citenamefont {Gebauer}, \citenamefont {Gerstmann}, \citenamefont {Giustino}, \citenamefont {Gorni}, \citenamefont {Jia}, \citenamefont {Kawamura}, \citenamefont {Ko}, \citenamefont {Kokalj}, \citenamefont {Küçükbenli}, \citenamefont {Lazzeri}, \citenamefont {Marsili}, \citenamefont {Marzari}, \citenamefont {Mauri}, \citenamefont {Nguyen}, \citenamefont {Nguyen}, \citenamefont {de-la Roza}, \citenamefont {Paulatto}, \citenamefont {Poncé}, \citenamefont
  {Rocca}, \citenamefont {Sabatini}, \citenamefont {Santra}, \citenamefont {Schlipf}, \citenamefont {Seitsonen}, \citenamefont {Smogunov}, \citenamefont {Timrov}, \citenamefont {Thonhauser}, \citenamefont {Umari}, \citenamefont {Vast}, \citenamefont {Wu},\ and\ \citenamefont {Baroni}}]{QE-2017}%
  \BibitemOpen
  \bibfield  {author} {\bibinfo {author} {\bibfnamefont {P.}~\bibnamefont {Giannozzi}}, \bibinfo {author} {\bibfnamefont {O.}~\bibnamefont {Andreussi}}, \bibinfo {author} {\bibfnamefont {T.}~\bibnamefont {Brumme}}, \bibinfo {author} {\bibfnamefont {O.}~\bibnamefont {Bunau}}, \bibinfo {author} {\bibfnamefont {M.~B.}\ \bibnamefont {Nardelli}}, \bibinfo {author} {\bibfnamefont {M.}~\bibnamefont {Calandra}}, \bibinfo {author} {\bibfnamefont {R.}~\bibnamefont {Car}}, \bibinfo {author} {\bibfnamefont {C.}~\bibnamefont {Cavazzoni}}, \bibinfo {author} {\bibfnamefont {D.}~\bibnamefont {Ceresoli}}, \bibinfo {author} {\bibfnamefont {M.}~\bibnamefont {Cococcioni}}, \bibinfo {author} {\bibfnamefont {N.}~\bibnamefont {Colonna}}, \bibinfo {author} {\bibfnamefont {I.}~\bibnamefont {Carnimeo}}, \bibinfo {author} {\bibfnamefont {A.~D.}\ \bibnamefont {Corso}}, \bibinfo {author} {\bibfnamefont {S.}~\bibnamefont {de~Gironcoli}}, \bibinfo {author} {\bibfnamefont {P.}~\bibnamefont {Delugas}}, \bibinfo {author} {\bibfnamefont
  {R.~A.~D.}\ \bibnamefont {Jr}}, \bibinfo {author} {\bibfnamefont {A.}~\bibnamefont {Ferretti}}, \bibinfo {author} {\bibfnamefont {A.}~\bibnamefont {Floris}}, \bibinfo {author} {\bibfnamefont {G.}~\bibnamefont {Fratesi}}, \bibinfo {author} {\bibfnamefont {G.}~\bibnamefont {Fugallo}}, \bibinfo {author} {\bibfnamefont {R.}~\bibnamefont {Gebauer}}, \bibinfo {author} {\bibfnamefont {U.}~\bibnamefont {Gerstmann}}, \bibinfo {author} {\bibfnamefont {F.}~\bibnamefont {Giustino}}, \bibinfo {author} {\bibfnamefont {T.}~\bibnamefont {Gorni}}, \bibinfo {author} {\bibfnamefont {J.}~\bibnamefont {Jia}}, \bibinfo {author} {\bibfnamefont {M.}~\bibnamefont {Kawamura}}, \bibinfo {author} {\bibfnamefont {H.-Y.}\ \bibnamefont {Ko}}, \bibinfo {author} {\bibfnamefont {A.}~\bibnamefont {Kokalj}}, \bibinfo {author} {\bibfnamefont {E.}~\bibnamefont {Küçükbenli}}, \bibinfo {author} {\bibfnamefont {M.}~\bibnamefont {Lazzeri}}, \bibinfo {author} {\bibfnamefont {M.}~\bibnamefont {Marsili}}, \bibinfo {author} {\bibfnamefont
  {N.}~\bibnamefont {Marzari}}, \bibinfo {author} {\bibfnamefont {F.}~\bibnamefont {Mauri}}, \bibinfo {author} {\bibfnamefont {N.~L.}\ \bibnamefont {Nguyen}}, \bibinfo {author} {\bibfnamefont {H.-V.}\ \bibnamefont {Nguyen}}, \bibinfo {author} {\bibfnamefont {A.~O.}\ \bibnamefont {de-la Roza}}, \bibinfo {author} {\bibfnamefont {L.}~\bibnamefont {Paulatto}}, \bibinfo {author} {\bibfnamefont {S.}~\bibnamefont {Poncé}}, \bibinfo {author} {\bibfnamefont {D.}~\bibnamefont {Rocca}}, \bibinfo {author} {\bibfnamefont {R.}~\bibnamefont {Sabatini}}, \bibinfo {author} {\bibfnamefont {B.}~\bibnamefont {Santra}}, \bibinfo {author} {\bibfnamefont {M.}~\bibnamefont {Schlipf}}, \bibinfo {author} {\bibfnamefont {A.~P.}\ \bibnamefont {Seitsonen}}, \bibinfo {author} {\bibfnamefont {A.}~\bibnamefont {Smogunov}}, \bibinfo {author} {\bibfnamefont {I.}~\bibnamefont {Timrov}}, \bibinfo {author} {\bibfnamefont {T.}~\bibnamefont {Thonhauser}}, \bibinfo {author} {\bibfnamefont {P.}~\bibnamefont {Umari}}, \bibinfo {author}
  {\bibfnamefont {N.}~\bibnamefont {Vast}}, \bibinfo {author} {\bibfnamefont {X.}~\bibnamefont {Wu}},\ and\ \bibinfo {author} {\bibfnamefont {S.}~\bibnamefont {Baroni}},\ }\bibfield  {title} {\bibinfo {title} {Advanced capabilities for materials modelling with quantum espresso},\ }\href {http://stacks.iop.org/0953-8984/29/i=46/a=465901} {\bibfield  {journal} {\bibinfo  {journal} {Journal of Physics: Condensed Matter}\ }\textbf {\bibinfo {volume} {29}},\ \bibinfo {pages} {465901} (\bibinfo {year} {2017})}\BibitemShut {NoStop}%
\bibitem [{\citenamefont {Box}\ \emph {et~al.}(2023)\citenamefont {Box}, \citenamefont {Stark},\ and\ \citenamefont {Maurer}}]{Box2023-fhiaims-fric}%
  \BibitemOpen
  \bibfield  {author} {\bibinfo {author} {\bibfnamefont {C.~L.}\ \bibnamefont {Box}}, \bibinfo {author} {\bibfnamefont {W.~G.}\ \bibnamefont {Stark}},\ and\ \bibinfo {author} {\bibfnamefont {R.~J.}\ \bibnamefont {Maurer}},\ }\bibfield  {title} {\bibinfo {title} {Ab initio calculation of electron-phonon linewidths and molecular dynamics with electronic friction at metal surfaces with numeric atom-centred orbitals},\ }\href {https://doi.org/10.1088/2516-1075/acf3c4} {\bibfield  {journal} {\bibinfo  {journal} {Electronic Structure}\ }\textbf {\bibinfo {volume} {5}},\ \bibinfo {pages} {035005} (\bibinfo {year} {2023})}\BibitemShut {NoStop}%
\bibitem [{\citenamefont {Bianco}\ \emph {et~al.}(2017)\citenamefont {Bianco}, \citenamefont {Errea}, \citenamefont {Paulatto}, \citenamefont {Calandra},\ and\ \citenamefont {Mauri}}]{Bianco2017-SSCHA1}%
  \BibitemOpen
  \bibfield  {author} {\bibinfo {author} {\bibfnamefont {R.}~\bibnamefont {Bianco}}, \bibinfo {author} {\bibfnamefont {I.}~\bibnamefont {Errea}}, \bibinfo {author} {\bibfnamefont {L.}~\bibnamefont {Paulatto}}, \bibinfo {author} {\bibfnamefont {M.}~\bibnamefont {Calandra}},\ and\ \bibinfo {author} {\bibfnamefont {F.}~\bibnamefont {Mauri}},\ }\bibfield  {title} {\bibinfo {title} {Second-order structural phase transitions, free energy curvature, and temperature-dependent anharmonic phonons in the self-consistent harmonic approximation: Theory and stochastic implementation},\ }\bibfield  {journal} {\bibinfo  {journal} {Physical Review B}\ }\textbf {\bibinfo {volume} {96}},\ \href {https://doi.org/10.1103/physrevb.96.014111} {10.1103/physrevb.96.014111} (\bibinfo {year} {2017})\BibitemShut {NoStop}%
\bibitem [{\citenamefont {Monacelli}\ \emph {et~al.}(2018)\citenamefont {Monacelli}, \citenamefont {Errea}, \citenamefont {Calandra},\ and\ \citenamefont {Mauri}}]{Monacelli2018-SSCHA2}%
  \BibitemOpen
  \bibfield  {author} {\bibinfo {author} {\bibfnamefont {L.}~\bibnamefont {Monacelli}}, \bibinfo {author} {\bibfnamefont {I.}~\bibnamefont {Errea}}, \bibinfo {author} {\bibfnamefont {M.}~\bibnamefont {Calandra}},\ and\ \bibinfo {author} {\bibfnamefont {F.}~\bibnamefont {Mauri}},\ }\bibfield  {title} {\bibinfo {title} {Pressure and stress tensor of complex anharmonic crystals within the stochastic self-consistent harmonic approximation},\ }\bibfield  {journal} {\bibinfo  {journal} {Physical Review B}\ }\textbf {\bibinfo {volume} {98}},\ \href {https://doi.org/10.1103/physrevb.98.024106} {10.1103/physrevb.98.024106} (\bibinfo {year} {2018})\BibitemShut {NoStop}%
\bibitem [{\citenamefont {Monacelli}\ \emph {et~al.}(2021)\citenamefont {Monacelli}, \citenamefont {Bianco}, \citenamefont {Cherubini}, \citenamefont {Calandra}, \citenamefont {Errea},\ and\ \citenamefont {Mauri}}]{Monacelli2021-SSCHA3}%
  \BibitemOpen
  \bibfield  {author} {\bibinfo {author} {\bibfnamefont {L.}~\bibnamefont {Monacelli}}, \bibinfo {author} {\bibfnamefont {R.}~\bibnamefont {Bianco}}, \bibinfo {author} {\bibfnamefont {M.}~\bibnamefont {Cherubini}}, \bibinfo {author} {\bibfnamefont {M.}~\bibnamefont {Calandra}}, \bibinfo {author} {\bibfnamefont {I.}~\bibnamefont {Errea}},\ and\ \bibinfo {author} {\bibfnamefont {F.}~\bibnamefont {Mauri}},\ }\bibfield  {title} {\bibinfo {title} {The stochastic self-consistent harmonic approximation: calculating vibrational properties of materials with full quantum and anharmonic effects},\ }\href {https://doi.org/10.1088/1361-648x/ac066b} {\bibfield  {journal} {\bibinfo  {journal} {Journal of Physics: Condensed Matter}\ }\textbf {\bibinfo {volume} {33}},\ \bibinfo {pages} {363001} (\bibinfo {year} {2021})}\BibitemShut {NoStop}%
\bibitem [{\citenamefont {Gao}\ \emph {et~al.}(1996)\citenamefont {Gao}, \citenamefont {Carr}, \citenamefont {Porter}, \citenamefont {Tanner}, \citenamefont {Williams}, \citenamefont {Hirschmugl}, \citenamefont {Dutta}, \citenamefont {Wu},\ and\ \citenamefont {Etemad}}]{PhysRevB.54.700}%
  \BibitemOpen
  \bibfield  {author} {\bibinfo {author} {\bibfnamefont {F.}~\bibnamefont {Gao}}, \bibinfo {author} {\bibfnamefont {G.~L.}\ \bibnamefont {Carr}}, \bibinfo {author} {\bibfnamefont {C.~D.}\ \bibnamefont {Porter}}, \bibinfo {author} {\bibfnamefont {D.~B.}\ \bibnamefont {Tanner}}, \bibinfo {author} {\bibfnamefont {G.~P.}\ \bibnamefont {Williams}}, \bibinfo {author} {\bibfnamefont {C.~J.}\ \bibnamefont {Hirschmugl}}, \bibinfo {author} {\bibfnamefont {B.}~\bibnamefont {Dutta}}, \bibinfo {author} {\bibfnamefont {X.~D.}\ \bibnamefont {Wu}},\ and\ \bibinfo {author} {\bibfnamefont {S.}~\bibnamefont {Etemad}},\ }\bibfield  {title} {\bibinfo {title} {Quasiparticle damping and the coherence peak in ${\mathrm{yba}}_{2}$${\mathrm{cu}}_{3}$${\mathrm{o}}_{7\mathrm{\ensuremath{-}}\mathrm{\ensuremath{\delta}}}$},\ }\href {https://doi.org/10.1103/PhysRevB.54.700} {\bibfield  {journal} {\bibinfo  {journal} {Phys. Rev. B}\ }\textbf {\bibinfo {volume} {54}},\ \bibinfo {pages} {700} (\bibinfo {year} {1996})}\BibitemShut {NoStop}%
\bibitem [{\citenamefont {Blum}\ \emph {et~al.}(2009)\citenamefont {Blum}, \citenamefont {Gehrke}, \citenamefont {Hanke}, \citenamefont {Havu}, \citenamefont {Havu}, \citenamefont {Ren}, \citenamefont {Reuter},\ and\ \citenamefont {Scheffler}}]{Blum2009-fhiaims}%
  \BibitemOpen
  \bibfield  {author} {\bibinfo {author} {\bibfnamefont {V.}~\bibnamefont {Blum}}, \bibinfo {author} {\bibfnamefont {R.}~\bibnamefont {Gehrke}}, \bibinfo {author} {\bibfnamefont {F.}~\bibnamefont {Hanke}}, \bibinfo {author} {\bibfnamefont {P.}~\bibnamefont {Havu}}, \bibinfo {author} {\bibfnamefont {V.}~\bibnamefont {Havu}}, \bibinfo {author} {\bibfnamefont {X.}~\bibnamefont {Ren}}, \bibinfo {author} {\bibfnamefont {K.}~\bibnamefont {Reuter}},\ and\ \bibinfo {author} {\bibfnamefont {M.}~\bibnamefont {Scheffler}},\ }\bibfield  {title} {\bibinfo {title} {Ab initio molecular simulations with numeric atom-centered orbitals},\ }\href {https://doi.org/10.1016/j.cpc.2009.06.022} {\bibfield  {journal} {\bibinfo  {journal} {Computer Physics Communications}\ }\textbf {\bibinfo {volume} {180}},\ \bibinfo {pages} {2175–2196} (\bibinfo {year} {2009})}\BibitemShut {NoStop}%
\bibitem [{\citenamefont {Eliashberg}(1972)}]{eliashberg1972inelastic}%
  \BibitemOpen
  \bibfield  {author} {\bibinfo {author} {\bibfnamefont {G.}~\bibnamefont {Eliashberg}},\ }\bibfield  {title} {\bibinfo {title} {Inelastic electron collisions and nonequilibrium stationary states in superconductors},\ }\href@noop {} {\bibfield  {journal} {\bibinfo  {journal} {Sov Phys JETP}\ }\textbf {\bibinfo {volume} {34}},\ \bibinfo {pages} {668} (\bibinfo {year} {1972})}\BibitemShut {NoStop}%
\bibitem [{\citenamefont {Akashi}\ and\ \citenamefont {Arita}(2013)}]{K3C60-Arita-2013}%
  \BibitemOpen
  \bibfield  {author} {\bibinfo {author} {\bibfnamefont {R.}~\bibnamefont {Akashi}}\ and\ \bibinfo {author} {\bibfnamefont {R.}~\bibnamefont {Arita}},\ }\bibfield  {title} {\bibinfo {title} {Nonempirical study of superconductivity in alkali-doped fullerides based on density functional theory for superconductors},\ }\href {https://doi.org/10.1103/PhysRevB.88.054510} {\bibfield  {journal} {\bibinfo  {journal} {Phys. Rev. B}\ }\textbf {\bibinfo {volume} {88}},\ \bibinfo {pages} {054510} (\bibinfo {year} {2013})}\BibitemShut {NoStop}%
\bibitem [{\citenamefont {Nomura}\ \emph {et~al.}(2015)\citenamefont {Nomura}, \citenamefont {Sakai}, \citenamefont {Capone},\ and\ \citenamefont {Arita}}]{doi:10.1126/sciadv.1500568}%
  \BibitemOpen
  \bibfield  {author} {\bibinfo {author} {\bibfnamefont {Y.}~\bibnamefont {Nomura}}, \bibinfo {author} {\bibfnamefont {S.}~\bibnamefont {Sakai}}, \bibinfo {author} {\bibfnamefont {M.}~\bibnamefont {Capone}},\ and\ \bibinfo {author} {\bibfnamefont {R.}~\bibnamefont {Arita}},\ }\bibfield  {title} {\bibinfo {title} {Unified understanding of superconductivity and mott transition in alkali-doped fullerides from first principles},\ }\href {https://doi.org/10.1126/sciadv.1500568} {\bibfield  {journal} {\bibinfo  {journal} {Science Advances}\ }\textbf {\bibinfo {volume} {1}},\ \bibinfo {pages} {e1500568} (\bibinfo {year} {2015})},\ \Eprint {https://arxiv.org/abs/https://www.science.org/doi/pdf/10.1126/sciadv.1500568} {https://www.science.org/doi/pdf/10.1126/sciadv.1500568} \BibitemShut {NoStop}%
\bibitem [{\citenamefont {Perdew}\ \emph {et~al.}(1996)\citenamefont {Perdew}, \citenamefont {Burke},\ and\ \citenamefont {Ernzerhof}}]{perdew_generalized_PBE_1996}%
  \BibitemOpen
  \bibfield  {author} {\bibinfo {author} {\bibfnamefont {J.~P.}\ \bibnamefont {Perdew}}, \bibinfo {author} {\bibfnamefont {K.}~\bibnamefont {Burke}},\ and\ \bibinfo {author} {\bibfnamefont {M.}~\bibnamefont {Ernzerhof}},\ }\bibfield  {title} {\bibinfo {title} {Generalized gradient approximation made simple},\ }\href {https://doi.org/10.1103/PhysRevLett.77.3865} {\bibfield  {journal} {\bibinfo  {journal} {Phys. Rev. Lett.}\ }\textbf {\bibinfo {volume} {77}},\ \bibinfo {pages} {3865} (\bibinfo {year} {1996})}\BibitemShut {NoStop}%
\bibitem [{\citenamefont {Methfessel}\ and\ \citenamefont {Paxton}(1989)}]{Methfessel_PRB_1989_smearing}%
  \BibitemOpen
  \bibfield  {author} {\bibinfo {author} {\bibfnamefont {M.}~\bibnamefont {Methfessel}}\ and\ \bibinfo {author} {\bibfnamefont {A.}~\bibnamefont {Paxton}},\ }\bibfield  {title} {\bibinfo {title} {High-precision sampling for brillouin-zone integration in metals},\ }\href@noop {} {\bibfield  {journal} {\bibinfo  {journal} {physical review B}\ }\textbf {\bibinfo {volume} {40}},\ \bibinfo {pages} {3616} (\bibinfo {year} {1989})}\BibitemShut {NoStop}%
\bibitem [{\citenamefont {Stephens}\ \emph {et~al.}(1991)\citenamefont {Stephens}, \citenamefont {Mihaly}, \citenamefont {Lee}, \citenamefont {Whetten}, \citenamefont {Huang}, \citenamefont {Kaner}, \citenamefont {Deiderich},\ and\ \citenamefont {Holczer}}]{Stephens1991}%
  \BibitemOpen
  \bibfield  {author} {\bibinfo {author} {\bibfnamefont {P.~W.}\ \bibnamefont {Stephens}}, \bibinfo {author} {\bibfnamefont {L.}~\bibnamefont {Mihaly}}, \bibinfo {author} {\bibfnamefont {P.~L.}\ \bibnamefont {Lee}}, \bibinfo {author} {\bibfnamefont {R.~L.}\ \bibnamefont {Whetten}}, \bibinfo {author} {\bibfnamefont {S.-M.}\ \bibnamefont {Huang}}, \bibinfo {author} {\bibfnamefont {R.}~\bibnamefont {Kaner}}, \bibinfo {author} {\bibfnamefont {F.}~\bibnamefont {Deiderich}},\ and\ \bibinfo {author} {\bibfnamefont {K.}~\bibnamefont {Holczer}},\ }\bibfield  {title} {\bibinfo {title} {Structure of single-phase superconducting k3c60},\ }\href {https://doi.org/10.1038/351632a0} {\bibfield  {journal} {\bibinfo  {journal} {Nature}\ }\textbf {\bibinfo {volume} {351}},\ \bibinfo {pages} {632–634} (\bibinfo {year} {1991})}\BibitemShut {NoStop}%
\bibitem [{\citenamefont {Giannozzi}\ \emph {et~al.}(2020)\citenamefont {Giannozzi}, \citenamefont {Baseggio}, \citenamefont {Bonfà}, \citenamefont {Brunato}, \citenamefont {Car}, \citenamefont {Carnimeo}, \citenamefont {Cavazzoni}, \citenamefont {de~Gironcoli}, \citenamefont {Delugas}, \citenamefont {Ferrari~Ruffino}, \citenamefont {Ferretti}, \citenamefont {Marzari}, \citenamefont {Timrov}, \citenamefont {Urru},\ and\ \citenamefont {Baroni}}]{QE-2020}%
  \BibitemOpen
  \bibfield  {author} {\bibinfo {author} {\bibfnamefont {P.}~\bibnamefont {Giannozzi}}, \bibinfo {author} {\bibfnamefont {O.}~\bibnamefont {Baseggio}}, \bibinfo {author} {\bibfnamefont {P.}~\bibnamefont {Bonfà}}, \bibinfo {author} {\bibfnamefont {D.}~\bibnamefont {Brunato}}, \bibinfo {author} {\bibfnamefont {R.}~\bibnamefont {Car}}, \bibinfo {author} {\bibfnamefont {I.}~\bibnamefont {Carnimeo}}, \bibinfo {author} {\bibfnamefont {C.}~\bibnamefont {Cavazzoni}}, \bibinfo {author} {\bibfnamefont {S.}~\bibnamefont {de~Gironcoli}}, \bibinfo {author} {\bibfnamefont {P.}~\bibnamefont {Delugas}}, \bibinfo {author} {\bibfnamefont {F.}~\bibnamefont {Ferrari~Ruffino}}, \bibinfo {author} {\bibfnamefont {A.}~\bibnamefont {Ferretti}}, \bibinfo {author} {\bibfnamefont {N.}~\bibnamefont {Marzari}}, \bibinfo {author} {\bibfnamefont {I.}~\bibnamefont {Timrov}}, \bibinfo {author} {\bibfnamefont {A.}~\bibnamefont {Urru}},\ and\ \bibinfo {author} {\bibfnamefont {S.}~\bibnamefont {Baroni}},\ }\bibfield  {title} {\bibinfo {title}
  {Quantum espresso toward the exascale},\ }\href {https://doi.org/10.1063/5.0005082} {\bibfield  {journal} {\bibinfo  {journal} {The Journal of Chemical Physics}\ }\textbf {\bibinfo {volume} {152}},\ \bibinfo {pages} {154105} (\bibinfo {year} {2020})}\BibitemShut {NoStop}%
\bibitem [{\citenamefont {Hamann}(2013{\natexlab{a}})}]{vanbilt_pseudo_hamann_2013_PhysRevB.88.085117}%
  \BibitemOpen
  \bibfield  {author} {\bibinfo {author} {\bibfnamefont {D.~R.}\ \bibnamefont {Hamann}},\ }\bibfield  {title} {\bibinfo {title} {Optimized norm-conserving vanderbilt pseudopotentials},\ }\href {https://doi.org/10.1103/PhysRevB.88.085117} {\bibfield  {journal} {\bibinfo  {journal} {Phys. Rev. B}\ }\textbf {\bibinfo {volume} {88}},\ \bibinfo {pages} {085117} (\bibinfo {year} {2013}{\natexlab{a}})}\BibitemShut {NoStop}%
\bibitem [{\citenamefont {Errea}\ \emph {et~al.}(2020)\citenamefont {Errea}, \citenamefont {Belli}, \citenamefont {Monacelli}, \citenamefont {Sanna}, \citenamefont {Koretsune}, \citenamefont {Tadano}, \citenamefont {Bianco}, \citenamefont {Calandra}, \citenamefont {Arita}, \citenamefont {Mauri},\ and\ \citenamefont {Flores-Livas}}]{Errea2020-SSCHA-LaH10}%
  \BibitemOpen
  \bibfield  {author} {\bibinfo {author} {\bibfnamefont {I.}~\bibnamefont {Errea}}, \bibinfo {author} {\bibfnamefont {F.}~\bibnamefont {Belli}}, \bibinfo {author} {\bibfnamefont {L.}~\bibnamefont {Monacelli}}, \bibinfo {author} {\bibfnamefont {A.}~\bibnamefont {Sanna}}, \bibinfo {author} {\bibfnamefont {T.}~\bibnamefont {Koretsune}}, \bibinfo {author} {\bibfnamefont {T.}~\bibnamefont {Tadano}}, \bibinfo {author} {\bibfnamefont {R.}~\bibnamefont {Bianco}}, \bibinfo {author} {\bibfnamefont {M.}~\bibnamefont {Calandra}}, \bibinfo {author} {\bibfnamefont {R.}~\bibnamefont {Arita}}, \bibinfo {author} {\bibfnamefont {F.}~\bibnamefont {Mauri}},\ and\ \bibinfo {author} {\bibfnamefont {J.~A.}\ \bibnamefont {Flores-Livas}},\ }\bibfield  {title} {\bibinfo {title} {Quantum crystal structure in the 250-kelvin superconducting lanthanum hydride},\ }\href {https://doi.org/10.1038/s41586-020-1955-z} {\bibfield  {journal} {\bibinfo  {journal} {Nature}\ }\textbf {\bibinfo {volume} {578}},\ \bibinfo {pages} {66–69} (\bibinfo
  {year} {2020})}\BibitemShut {NoStop}%
\bibitem [{\citenamefont {Shapeev}(2016)}]{Shapeev2016-MTP1}%
  \BibitemOpen
  \bibfield  {author} {\bibinfo {author} {\bibfnamefont {A.~V.}\ \bibnamefont {Shapeev}},\ }\bibfield  {title} {\bibinfo {title} {Moment tensor potentials: A class of systematically improvable interatomic potentials},\ }\href {https://doi.org/10.1137/15m1054183} {\bibfield  {journal} {\bibinfo  {journal} {Multiscale Modeling and Simulation}\ }\textbf {\bibinfo {volume} {14}},\ \bibinfo {pages} {1153–1173} (\bibinfo {year} {2016})}\BibitemShut {NoStop}%
\bibitem [{\citenamefont {Novikov}\ \emph {et~al.}(2021)\citenamefont {Novikov}, \citenamefont {Gubaev}, \citenamefont {Podryabinkin},\ and\ \citenamefont {Shapeev}}]{Novikov2021-MTP2}%
  \BibitemOpen
  \bibfield  {author} {\bibinfo {author} {\bibfnamefont {I.~S.}\ \bibnamefont {Novikov}}, \bibinfo {author} {\bibfnamefont {K.}~\bibnamefont {Gubaev}}, \bibinfo {author} {\bibfnamefont {E.~V.}\ \bibnamefont {Podryabinkin}},\ and\ \bibinfo {author} {\bibfnamefont {A.~V.}\ \bibnamefont {Shapeev}},\ }\bibfield  {title} {\bibinfo {title} {The mlip package: moment tensor potentials with mpi and active learning},\ }\href {https://doi.org/10.1088/2632-2153/abc9fe} {\bibfield  {journal} {\bibinfo  {journal} {Machine Learning: Science and Technology}\ }\textbf {\bibinfo {volume} {2}},\ \bibinfo {pages} {025002} (\bibinfo {year} {2021})}\BibitemShut {NoStop}%
\bibitem [{\citenamefont {Lucrezi}\ \emph {et~al.}(2023)\citenamefont {Lucrezi}, \citenamefont {Kogler}, \citenamefont {Cataldo}, \citenamefont {Aichhorn}, \citenamefont {Boeri},\ and\ \citenamefont {Heil}}]{Lucrezi2023}%
  \BibitemOpen
  \bibfield  {author} {\bibinfo {author} {\bibfnamefont {R.}~\bibnamefont {Lucrezi}}, \bibinfo {author} {\bibfnamefont {E.}~\bibnamefont {Kogler}}, \bibinfo {author} {\bibfnamefont {S.}~\bibnamefont {Cataldo}}, \bibinfo {author} {\bibfnamefont {M.}~\bibnamefont {Aichhorn}}, \bibinfo {author} {\bibfnamefont {L.}~\bibnamefont {Boeri}},\ and\ \bibinfo {author} {\bibfnamefont {C.}~\bibnamefont {Heil}},\ }\bibfield  {title} {\bibinfo {title} {Quantum lattice dynamics and their importance in ternary superhydride clathrates},\ }\href {https://doi.org/10.1038/s42005-023-01413-8} {\bibfield  {journal} {\bibinfo  {journal} {Communications Physics}\ }\textbf {\bibinfo {volume} {6}},\ \bibinfo {pages} {298} (\bibinfo {year} {2023})}\BibitemShut {NoStop}%
\bibitem [{\citenamefont {Hamann}(2013{\natexlab{b}})}]{ONCV1}%
  \BibitemOpen
  \bibfield  {author} {\bibinfo {author} {\bibfnamefont {D.~R.}\ \bibnamefont {Hamann}},\ }\bibfield  {title} {\bibinfo {title} {Optimized norm-conserving vanderbilt pseudopotentials},\ }\href {https://doi.org/10.1103/PhysRevB.88.085117} {\bibfield  {journal} {\bibinfo  {journal} {Phys. Rev. B}\ }\textbf {\bibinfo {volume} {88}},\ \bibinfo {pages} {085117} (\bibinfo {year} {2013}{\natexlab{b}})}\BibitemShut {NoStop}%
\bibitem [{\citenamefont {Schlipf}\ and\ \citenamefont {Gygi}(2015)}]{ONCV2}%
  \BibitemOpen
  \bibfield  {author} {\bibinfo {author} {\bibfnamefont {M.}~\bibnamefont {Schlipf}}\ and\ \bibinfo {author} {\bibfnamefont {F.}~\bibnamefont {Gygi}},\ }\bibfield  {title} {\bibinfo {title} {Optimization algorithm for the generation of oncv pseudopotentials},\ }\href {https://doi.org/https://doi.org/10.1016/j.cpc.2015.05.011} {\bibfield  {journal} {\bibinfo  {journal} {Computer Physics Communications}\ }\textbf {\bibinfo {volume} {196}},\ \bibinfo {pages} {36} (\bibinfo {year} {2015})}\BibitemShut {NoStop}%
\bibitem [{\citenamefont {{van Setten}}\ \emph {et~al.}(2018)\citenamefont {{van Setten}}, \citenamefont {Giantomassi}, \citenamefont {Bousquet}, \citenamefont {Verstraete}, \citenamefont {Hamann}, \citenamefont {Gonze},\ and\ \citenamefont {Rignanese}}]{setten2018}%
  \BibitemOpen
  \bibfield  {author} {\bibinfo {author} {\bibfnamefont {M.}~\bibnamefont {{van Setten}}}, \bibinfo {author} {\bibfnamefont {M.}~\bibnamefont {Giantomassi}}, \bibinfo {author} {\bibfnamefont {E.}~\bibnamefont {Bousquet}}, \bibinfo {author} {\bibfnamefont {M.}~\bibnamefont {Verstraete}}, \bibinfo {author} {\bibfnamefont {D.}~\bibnamefont {Hamann}}, \bibinfo {author} {\bibfnamefont {X.}~\bibnamefont {Gonze}},\ and\ \bibinfo {author} {\bibfnamefont {G.-M.}\ \bibnamefont {Rignanese}},\ }\bibfield  {title} {\bibinfo {title} {The pseudodojo: Training and grading a 85 element optimized norm-conserving pseudopotential table},\ }\href {https://doi.org/https://doi.org/10.1016/j.cpc.2018.01.012} {\bibfield  {journal} {\bibinfo  {journal} {Computer Physics Communications}\ }\textbf {\bibinfo {volume} {226}},\ \bibinfo {pages} {39} (\bibinfo {year} {2018})}\BibitemShut {NoStop}%
\bibitem [{\citenamefont {Monkhorst}\ and\ \citenamefont {Pack}(1976)}]{monkhorst1976}%
  \BibitemOpen
  \bibfield  {author} {\bibinfo {author} {\bibfnamefont {H.~J.}\ \bibnamefont {Monkhorst}}\ and\ \bibinfo {author} {\bibfnamefont {J.~D.}\ \bibnamefont {Pack}},\ }\bibfield  {title} {\bibinfo {title} {Special points for brillouin-zone integrations},\ }\href {https://doi.org/10.1103/PhysRevB.13.5188} {\bibfield  {journal} {\bibinfo  {journal} {Phys. Rev. B}\ }\textbf {\bibinfo {volume} {13}},\ \bibinfo {pages} {5188} (\bibinfo {year} {1976})}\BibitemShut {NoStop}%
\end{thebibliography}
\end{document}